\documentclass[prb,twocolumn,superscriptaddress,showpacs,preprintnumbers,amsmath,amssymb,twoside]{revtex4}

\usepackage{latexsym,amsmath,graphics,graphicx}
\newcommand{\lbl}{\label}
\newcommand{\sgm}{\boldsymbol{\sigma}}
\newcommand{\vc}{\mathbf}

\begin{document}
\title{A KKR Green function formalism for ballistic transport}
\author{Phivos Mavropoulos}
\affiliation{Institut f\"ur Festk\"orperforschung, Forschungszentrum
J\"ulich, D-52425 J\"ulich, Germany}
\author{Nikolaos Papanikolaou}
\affiliation{Institute of Microelectronics, NCSR ``Democritos'',
  GR-15310 Ag.~Paraskevi, Attiki, Greece}
\author{Peter H.~Dederichs}
\affiliation{Institut f\"ur Festk\"orperforschung, Forschungszentrum
J\"ulich, D-52425 J\"ulich, Germany}
\date{\today}

%%%%%%%%%%%%%%%%%%%%%%%%%%%%%%%%%%%%%%%%%%%%%%%%%%%%%%%%%%%%%%%%%%%%%%
\begin{abstract}
We develop a method for the calculation of ballistic transport from
first principles. The multiple scattering screened
Korringa-Kohn-Rostoker (KKR) method is combined with a Green's
function formulation of the Landauer approach for the ballistic
transport. We obtain an efficient O(N) algorithm for the calculation
of ballistic conductance through a scattering region connected to
semi-infinite crystalline leads. In particular we generalize the
results of Baranger and Stone in the case of Bloch wave boundary
conditions and we discuss relevant properties of the S-matrix. We
consider the implications on the application of the formalism in
conjunction with a cellular multiple scattering description of the
electronic structure, and demonstrate the convergence properties
concerning the angular momentum expansions.
\end{abstract}
%%%%%%%%%%%%%%%%%%%%%%%%%%%%%%%%%%%%%%%%%%%%%%%%%%%%%%%%%%%%%%%%%%%%%%
\pacs{72.10.Bg,72.15.-v,71.15.-m}
%72.      Electronic transport in condensed matter
%72.10.-d Theory of electronic transport; scattering mechanisms
%72.10.Bg General formulation of transport theory
%72.15.-v Electronic conduction in metals and alloys
%72.15.Eb Electrical and thermal conduction in crystalline metals and alloys
%72.25.-b Spin polarized transport
%71.15.-m Methods of electronic structure calculations 

\maketitle
\section{Introduction}

The study of transport properties of solids has always been a
challenge for experimental and theoretical condensed matter physics.
The difficulty lies among other things in the fact that the transport
properties are by definition related to non-equilibrium situations, so
that their modeling is not easily based on standard techniques.
Fortunately, in the regime of linear response to weak external fields,
the study can be based on ground state properties, by treating the
field as a perturbation. That this approach is founded solidly is
guaranteed by the fluctuation-dissipation theorem; in particular for
the case of electrical conductivity this is expressed by the famous
result of Kubo.\cite{Kubo57}

%For the theoretical approach and calculation of the electronic
%transport, there exist many approaches, depending also on the nature
%of the system to be studied. Basically one can make the distinction
%between diffusive and ballistic transport. In the former case one is
%faced with the situation when the electron wavepacket is randomly
%de-phased during its passing through the material, due to inelastic or
%random scattering. Ballistic transport is defined as the opposite
%situation, where the wavepacket preserves fully the memory of its
%past, and where diffraction and interference effects are of
%importance. The area of validity of the two cases depends strongly on
%the temperature and the size of the sample under consideration. High
%temperatures result in inelastic electron scattering and pose us in
%the diffusive regime. As the temperature is lowered, ballistic effects
%become more and more important when the size of interest is small
%enough that no inelastic or other phase-randomizing events appear;
%this critical size is usually addressed as the mean free path.
%Various theoretical approaches exist for the understanding of
%electronic transport. For instance, diffusive transport can be
%confronted by the Boltzmann equation, the Keldysh method, or the
%Kubo-Greenwood approach; ballistic transport is often dealt with in
%the framework of Landauer-B\"uttiker theory or, again, the
%Kubo-Greenwood approach. 

In the past decades, the development of first-principles methods for
the calculation of the electronic structure of solids has been
accompanied by corresponding advances in the formalism and methods for
the calculation of transport properties. Technological interest has
given a push to the field, and novel effects such as the Giant and the
Tunneling Magnetoresistance (GMR and TMR) have been investigated
theoretically and applied in technology.  Furthermore, the size of
today's electronic devices is so small that the understanding of
ballistic ({\it i.e.}~phase-coherent) transport has become important
not only for basic physics but also for applications. As a result of
these developments, ideas and methods which initially were conceived
for the understanding of electronic transport in simple cases, have
been combined with techniques based on a realistic description of the
electronic structure in order to give reliable and material-specific
results.\cite{Weinberger03} Without claiming to give a complete list,
we mention that for diffusive transport through disordered systems
there exist methods and results based on the combination of the
Korringa, Kohn and Rostoker (KKR) Green function method with the
Boltzmann formalism,\cite{Mertig93,Mertig99,Zahn98} on the coherent
potential approximation (CPA) combined with
Kubo-Greenwood\cite{Kubo57,Greenwood58}
theory,\cite{Butler85,Banhart94,Weinberger96} or on the tight-binding
method.\cite{Turek02} For ballistic transport there exist methods
based, for example, on LEED,\cite{Pendry74,Modinos84}
layer-KKR\cite{MacLaren99} or similar layer-type\cite{Stiles88}
techniques, on the tight-binding
approach,\cite{Mathon97,Tsymbal01,Kristic02} and on the transfer
matrix concept;\cite{Wortmann02a,Wortmann02b} these mostly combine the
Landauer-B\"uttiker approach\cite{Landauer88,Buttiker86} with
electronic structure methods, as will be done in the present paper.
Moreover, due to potential applications in GMR and TMR devices,
spin-dependent transport has come to the center of interest with
emphasis to conduction in magnetic
multilayers\cite{Camblong93,Butler95,Itoh95,Weinberger96,Mathon97,Zahn98,Weinberger03}
and ferromagnet-semiconductor
hybrides.\cite{MacLaren99,Tsymbal01,Caroli71,Zhang99,Weinberger02,Wunnicke02,Mavropoulos02,Zwierzycki03}
In these systems the electronic spin degrees of freedom are accounted
for in order to achieve spin-dependent resistance. In addition, novel
systems such as nanowires or atomic-size contacts are created
experimentally, and demand interpretation of their transport
properties.

In this paper we present a method for the calculation of ballistic
transport from first principles, which combines the KKR {\it ab
  initio} Green function technique with the Baranger and
Stone\cite{Baranger89} formulation of the ballistic transport. The
method is suitable for layered systems and interfaces with
two-dimensional periodicity, as well as for atomic size constrictions
connecting two leads or nanowires,\cite{Ohnishi98} and supports spin
dependent effects. Since the KKR technique in the screened formalism
offers linear scaling of the computational effort with the number of
layers,\cite{Wildberger97} our method can handle large systems.
Additionally we present the proof of some theorems concerning the
transmission probability of wavepackets in crystalline, rather than
free-electron, environment.

The paper is organized as follows. In Section \ref{SecSetup} we give a
description of the setup of the physical systems that our method can
be used for. In Section \ref{SecKubo} we briefly address the
approximations and assumptions made, also in connection to the
Kubo-Greenwood and the Baranger-Stone formalism. Sections
\ref{SecLandauer} and \ref{SecKL} are devoted to making the connection
to the Landauer-B\"uttiker formalism, and some elements of scattering
theory are given there. The conductance formalism for KKR is developed
in Section \ref{SecKKR}. Some examples illustrating the convergence
properties of the method are given in Section \ref{SecExamples}.
Finally, we conclude in Section \ref{SecSummary}. The appendices
contain parts where lengthy mathematical manipulations were needed.

%%%%%%%%%%%%%%%%%%%%%%%%%%%%%%%%%%%%%%%%%%%%%%%%%%%%%%%%%%%%%%%%%%%%%%
\section{Setup of the problem \lbl{SecSetup}}

%In the Baranger-Stone formulation, the sample is connected to an
%arbitrary number of perfect infinite conducting free-electron leads,
%each of them considered in waveguide geometry. In an {\it ab initio}
%calculation, one wishes to have a realistic band structure of the
%leads. On the other hand, the inclusion of more than two leads is an
%almost impossible task numerically; rather, one is confined in a
%two-lead geometry, in a system periodic in two dimensions (say $x$
%and $y$) and grown in the third ($z$), so that the translational
%symmetry in $x$ and $y$ can be exploited in order to reduce the
%computation time to realistic levels. In this sense the formalism
%presented here is a special case of the Baranger-Stone result, with
%only two leads and in the absence of a magnetic field, but some points
%of that article must be re-proven, taking Bloch-electron, rather than
%free-electron, wavefunctions in the leads; in particular, the
%connection of the formalism to the Landauer-B\"uttiker theory needs
%special attention.

The systems that we study consist of two half infinite perfect
crystalline leads, left (L) and right (R), attached to a slab which is
considered as the ``interaction'' region (I); schematically this is
represented in Fig.~\ref{figSetup}. The surfaces $S_{\mathrm{L}}$ and
$S_{\mathrm{R}}$ should be chosen far enough within the leads so that
any localized interface states have decayed and only Bloch states are
present there. The direction of growth is taken to be the $z$
direction.  We consider systems with a 2D periodicity in the xy plane
where the scattering region is embedded between two semi-infinite
crystalline leads. Moreover, summing up current contributions in real
space we can calculate the conductance in the presence of defects in
nanowires\cite{Opitz02,Papanikolaou02b} and we can even treat atomic
sized constrictions between infinite leads if the current flow is
localized in the constriction region which is usually the
case.\cite{Bagrets03} Thus we can simulate transport through small
molecules or a break junction geometry.
\begin{figure}
\begin{center}
\includegraphics[width=5cm]{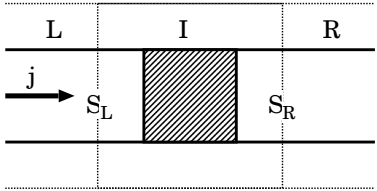}
\end{center}
\caption{Setup of the problem. The regions L and R correspond to
perfect semi-infinite crystalline leads (not necessarily of the same
material), attached to an ``interaction region'' I. This includes the
interface or other structures sandwiched between L and R, plus a few
monolayers of the leads at each side so that the evanescent interface
states can decay. The conductance is evaluated between the surfaces
$S_{\mathrm{L}}$ and $S_{\mathrm{R}}$, where only bulk Bloch states
should exist.}
\lbl{figSetup}
\end{figure}

%%%%%%%%%%%%%%%%%%%%%%%%%%%%%%%%%%%%%%%%%%%%%%%%%%%%%%%%%%%%%%%%%%%%%%
\section{Kubo formalism \lbl{SecKubo}}

The connection between the ballistic dc conductance or conductivity
and the one-electron Green function in the linear-response regime has
been given in the past.\cite{Fisher81,Stone88,Baranger89,Rammer91} In
short, one uses first-order perturbation theory to calculate the
effect of a weak, time-oscillating electric field on the density matrix
of the system; then, the frequency of the oscillating field is taken to
zero, and the dissipative term of the expectation value of the current
gives the dc linear response of the system. The result is just an
expression for the Kubo conductivity of the system at the limit of
zero temperature.  Integration by parts of the conductivity expression
gives the relation between current, conductance and voltage. Baranger
and Stone\cite{Baranger89} have proven that the conductance is a
Fermi-level property, although the two-point conductivity within the
sample can also have contributions from other energy levels when
magnetic fields are present. Here we briefly describe the relevant
formalism, in connection to the rest of the paper.

The general relation between the current density at some point,
$\mathbf{j}(\mathbf{r})$, and the electric field at all other points
of a material, $\mathbf{E}(\mathbf{r}')$, in first order perturbation
theory for a stationary state, is given by
\begin{equation}
\mathbf{j}(\mathbf{r}) = \int d^3r' \sgm(\mathbf{r},\mathbf{r}')
\mathbf{E}(\mathbf{r}') 
\lbl{eqOhm}
\end{equation}
The nonlocal conductivity tensor $\sgm(\mathbf{r},\mathbf{r}')$ is
connected to the retarded one-electron Green function
$G^+(\vc{r},\vc{r}';E)$ of the system via the famous Kubo-Greenwood
result. Its frequently used form is\cite{Butler85}
\begin{eqnarray}
  \sgm(\vc{r},\vc{r}') &=& -\frac{\hbar}{\pi}\int dE\, (-f'(E)) \lbl{eqKubo1}\\
&& \times\left(\frac{e\hbar}{m}\nabla \Im G^+(\vc{r},\vc{r}';E)\right) 
\left(\frac{e\hbar}{m}\nabla' \Im G^+(\vc{r}',\vc{r};E)\right)
\nonumber
\end{eqnarray}
Here, $f'(E)$ is the Fermi function derivative and $\Im$ denotes the
imaginary part. In terms of the difference between the retarded and
advanced Green function $G^-$
\begin{eqnarray}
\Delta G(\mathbf{r},\mathbf{r}';E) &:=&
G^+(\mathbf{r},\mathbf{r}';E) - G^-(\mathbf{r},\mathbf{r}';E) \\
&=& -2\pi i \int da \psi_a (\mathbf{r}) \psi_a^*(\mathbf{r}') \delta(E-E_a)
\end{eqnarray}
the conductivity can be written
\begin{eqnarray}
\sgm(\mathbf{r},\mathbf{r}') &=& - \frac{e^2 \hbar^3}{16\pi m^2}
\int dE (-f'(E)) \nonumber\\
&&\times\Delta G(\mathbf{r},\mathbf{r}';E)
\overleftrightarrow{\nabla}\overleftrightarrow{\nabla}'
\Delta G(\mathbf{r}',\mathbf{r};E)
\lbl{eqKubo}
\end{eqnarray}
where the symbol $\overleftrightarrow{\nabla}$ is defined as
$f(\vc{r}) \overleftrightarrow{\nabla} g(\vc{r}) = f(\vc{r}) \nabla
g(\vc{r}) - \left(\nabla f(\vc{r})\right) g(\vc{r})$.  Here only the
delta-function part of the Green function has been kept; the
principal-value part gives rise to the reactive term, which is zero
for dc conductance.\cite{Baranger89} Upon taking the limit of zero
temperature, the derivative of the Fermi-Dirac function becomes a
delta function according to $-f'(E)\rightarrow \delta(E-E_F)$ and
contributions to the conductivity only arise from the Fermi level.

In the absence of a magnetic field, as assumed in our work, the
nonlocal conductivity tensor is symmetric under the interchange of
$\vc{r}$ and $\vc{r}'$, and divergenceless:
\begin{eqnarray}
\sgm(\vc{r},\vc{r}') &=& \sgm(\vc{r}',\vc{r}) \\
\nabla\cdot\sgm(\vc{r},\vc{r}')&=&\sgm(\vc{r},\vc{r}')\cdot\overleftarrow{\nabla}'=0,
\lbl{eqDivergenceless}
\end{eqnarray}
where $\overleftarrow{\nabla}$ means that the operator acts to the
function on its left. The former relation is a consequence of the
symmetry $G(\vc{r},\vc{r}')=G(\vc{r}',\vc{r})$ of the Green function
for local potentials. The latter is a result of current conservation
in the absence of magnetic field; if a magnetic field is present, one
has merely
$\nabla\cdot\sgm(\vc{r},\vc{r}')\cdot\overleftarrow{\nabla}'=0$
instead.\cite{Baranger89}

In Eq.~(\ref{eqOhm}), of course, $\mathbf{E}(\mathbf{r}')$ is the
\emph{total} field, {\it i.e.}~the external one plus the one being
induced by charge relaxations; the electronic gas and nuclei must be
also fully taken into account. This is not easy to handle; however,
this expression can be integrated over the whole sample leaving only
the current in relation to the applied voltage by exploiting the fact
that $\mathbf{E}(\mathbf{r})=\nabla V(\mathbf{r})$, where
$V(\mathbf{r})$ is the electrostatic potential. Inserting this into
Eq.~(\ref{eqOhm}) one gets for the total current $J$ through
$S_{\mathrm{R}}$
\begin{equation}
J := \int_{S_{\mathrm{R}}} dS\,\hat{z} \,\mathbf{j}(\mathbf{r})
= \Delta V \int_{S_{\mathrm{R}}}
  dS\int_{S_{\mathrm{L}}} dS' \hat{z} \sgm(\mathbf{r},\mathbf{r}')\hat{z}',
\end{equation}
after an integration by parts and use of Eq.~(\ref{eqDivergenceless}). 
%Inserting this into the previous equation,
%integrating throughout the interaction region I of the sample, and
%making an integration by parts, one gets for the total current $J$
%through the cross-section $S_{\mathrm{R}}$ of the junction (see
%Fig.~\ref{figSetup}):
%\begin{eqnarray}
%  J &=& \int_{S_{\mathrm{R}}} dS\,\hat{z} \,\mathbf{j}(\mathbf{r}) \\ 
%&=& \int_{S_{\mathrm{R}}} dS \hat{z}
%  \int_{\mathrm{I}} d^3r' \sgm(\mathbf{r},\mathbf{r}') \nabla V(\mathbf{r}') \\ 
%  &=& \int_{S_{\mathrm{R}}} dS\hat{z}\int_{S(\mathrm{I})} dS'
%  \sgm(\mathbf{r},\mathbf{r}')\hat{n}' V(\mathbf{r}') \nonumber\\
%&&\mbox{} - \int_{S_{\mathrm{R}}}
%  dS\hat{z}\int_{S(\mathrm{I})} dS'
%  \sgm(\mathbf{r},\mathbf{r}')\overleftarrow{\nabla}'\hat{z} V(\mathbf{r}')
%\end{eqnarray}
%where $S(\mathrm{I})$ denotes the boundary of the region I and where an
%integration by parts has been used. The second term vanishes because
%the conductivity is divergenceless. 
Here $\Delta V$ is the external bias voltage.  The ``zero-current
theorem'' stating that the ground-state electrostatic potential gives
always zero current density\cite{Todorov93} has been used, so only the
externally applied voltage $\Delta V$ remains. Thus one can recognize
the conductance as the flux of the conductivity tensor through the
surfaces $S_{\mathrm{L}}$ and $S_{\mathrm{R}}$:
\begin{equation}
  g = \int_{S_{\mathrm{L}}} dS\int_{S_{\mathrm{R}}} dS' 
\hat{z}\sgm(\mathbf{r},\mathbf{r}')\hat{z}'
\end{equation}
Substituting Eq.~(\ref{eqKubo}) one obtains for the conductivity:
\begin{eqnarray}
g &=& - \frac{e^2 \hbar^3}{16\pi m^2} \int_{S_{\mathrm{L}}} dS\int_{S_{\mathrm{R}}} dS'
\nonumber\\
&&\times
G^+(\mathbf{r},\mathbf{r}';E_F) 
\overleftrightarrow{\nabla}_z \overleftrightarrow{\nabla}'_z
G^-(\mathbf{r}',\mathbf{r};E_F) 
\lbl{eqBaranger}
\end{eqnarray}
where the terms involving $G^+\overleftrightarrow{\nabla}_z
\overleftrightarrow{\nabla}'_z G^+$ and
$G^-\overleftrightarrow{\nabla}_z \overleftrightarrow{\nabla}'_z G^-$
vanish.\cite{Baranger89}

%%%%%%%%%%%%%%%%%%%%%%%%%%%%%%%%%%%%%%%%%%%%%%%%%%%%%%%%%%%%%%%%%%%%%%
\section{Landauer formalism \lbl{SecLandauer}}

In the Landauer approach, the conductance problem is viewed from the
aspect of scattering theory. In this way, the causal relation between
voltage and current is conceptually reversed.\cite{Imry99} Instead of
applying a voltage and examining the current as the response, a
current is forced to flow through the sample and the voltage is viewed
as the result of the pileup of carriers at the various obstacles,
forming residual resistivity dipoles. The result is, of course, again
the usual current-voltage relation. In a multi-lead experiment, if
each lead $n$ is held in potential $V_n$, then the current $J_n$ going
out through this lead is
\begin{equation}
J_n = \sum_{m\neq n} g_{nm}\, V_m.
\end{equation}
The coefficients $g_{nm}$ describe the conductance of the system, and
in the case of only two leads there is only one conductance
coefficient $g$.

The concept of incoming and outgoing \emph{scattering channels} is
introduced, which play the same role as in- and out-states in
scattering theory; in the cases of our interest they are Bloch states
in the leads. We need to describe the scattering process of
one-electron Bloch states incident from L as incoming waves and
scattered into L or R as outgoing waves; thus an $S$-matrix
formulation is appropriate. One has waveguide or crystalline geometry
in the leads, rather than the free-space geometry of usual scattering
theory, and in addition the leads can consist of different materials.
In this respect, usual scattering theory needs a few modifications to
be applicable. 

Once the $S$-matrix elements are available, one can readily calculate
the transmission probability $T_{nf;mi}$ from each incoming channel $i$
in lead $m$ to each outgoing $f$ in lead $n$, and use the
Landauer-B\"uttiker formula to calculate the conductance $g_{nm}$ from
lead $m$ to lead $n$:\cite{Buttiker86}
\begin{equation}
g_{nm} = \frac{e^2}{h}\sum_{fi} T_{nf;mi} 
= \frac{e^2}{h}\sum_{fi} |S_{nf;mi}|^2 
\end{equation}
Stone and Szafer\cite{Stone88} and Baranger and Stone\cite{Baranger89}
showed the connection between the Landauer-B\"uttiker formula with the
Kubo result starting from perturbation theory, considering
free-electron leads. In Sec.~\ref{SecKL} we shall follow their
analysis closely, but prove some of the theorems needed for crystalline
leads taking into account the Bloch character of the incoming and
outgoing channels.

The $S$-matrix formulation in such problems and the question of
unitarity will be shortly addressed now. First of all, the in-states
and out-states, which are in a usual formulation plane or spherical
waves, are here propagating Bloch states in the leads. Whether such a
state $\Psi_{\mathbf{k}}$ is incoming or outgoing is determined not by
the Bloch wavevector $\mathbf{k}$, but by the group velocity
$\mathbf{v}_{\mathbf{k}} = \nabla_{\mathbf{k}} E_{\mathbf{k}}$.
Suppose that a particular lead is grown in the $z$-direction, with the
unit vector $\hat{z}$ pointing away from the sample. Then, the Bloch
state is outgoing if $(\mathbf{v}_{\mathbf{k}})_z >0$, and incoming if
$(\mathbf{v}_{\mathbf{k}})_z < 0$. Evidently the set of in-states can
be different than the set of out-states. Consider, for example, the
case of only two leads of different materials, say L in the left with
wavefunctions $\Psi^{\mathrm{L}}$ and R in the right with
wavefunctions $\Psi^{\mathrm{R}}$, sandwiching the interaction region
I. The set of in-states for this problem will consist of
right-traveling waves $\Psi_{\mathrm{in}}^{\mathrm{L}}$ and
left-traveling $\Psi_{\mathrm{in}}^{\mathrm{R}}$, while the set of
out-states will consist of left-traveling waves
$\Psi_{\mathrm{out}}^{\mathrm{L}}$ and right-traveling waves
$\Psi_{\mathrm{out}}^{\mathrm{R}}$.

The total wavefunction in the system will be asymptotically
a linear combination of in- and out-states; in particular, if we
choose incidence from L as a boundary condition, we have the form
\begin{equation}
\Psi_{\mathbf{k}a}^{\mathrm{tot}}(\mathbf{r}) = \left\{
\begin{array}{ll}
\Psi_{\mathbf{k}a}^{\mathrm{L\, in}} +
\sum_{\mathbf{k}'a'} r_{\mathbf{k}a\mathbf{k}'a'} 
\Psi_{\mathbf{k}'a'}^{\mathrm{L\, out}} & \mbox{, $z\rightarrow -\infty$}
 \\ \\
\sum_{\mathbf{k}'a'} t_{\mathbf{k}a\mathbf{k}'a'} 
\Psi_{\mathbf{k}'a'}^{\mathrm{R\, out}} & \mbox{, $z\rightarrow +\infty$}
\end{array} \right.
\lbl{eqScattering}
\end{equation}
For finite $z$ values also evanescent states have to be included in
the summations, however they die out for $z\rightarrow\pm\infty$.  Here,
band indices $a$ and $a'$ and Bloch wavevectors $\mathbf{k}$ and
$\mathbf{k}'$, have been introduced; elastic scattering is implied.
The total wavefunction $\Psi_{\mathbf{k}a}^{\mathrm{tot}}(\mathbf{r})$
is characterized by $\mathbf{k}a$ in the sense of the boundary
condition, {\it i.e.}, it originates from an in-state
which has this wavevector.  The transmission amplitude
$t_{\mathbf{k}a\mathbf{k}'a'}$ and the reflection amplitude
$r_{\mathbf{k}a\mathbf{k}'a'}$ are related the elements of the
$S$-matrix of the system.

The normalization of the in- and out-states determines whether these
amplitudes $t$ and $r$ are identical with the $S$-matrix elements or
not, since the $S$-matrix must always be unitary. If one normalizes
the in- and out-states to unit flux (as $\Psi \rightarrow
\Psi/\sqrt{v}$), then the transmission probability from state
$\Psi_{\mathbf{k}a\,\mathrm{in}}^{\mathrm{L}}$ to state
$\Psi_{\mathbf{k}'a'\,\mathrm{out}}^{\mathrm{R}}$ is just
$T_{\mathrm{R}\,\mathbf{k}'a';\mathrm{L}\,\mathbf{k}a}=
|t_{\mathbf{k}a\mathbf{k}'a'}|^2= |S_{\mathbf{k}a\mathbf{k}'a'}|^2$.
But with the usual normalization to unit probability in space, the
transmission probability must invoke the group velocities to account
for the different flux of in- and out-states:\cite{footnote1}
\begin{equation}
T_{\mathrm{R}\,\mathbf{k}'a';\mathrm{L}\,\mathbf{k}a}=
  |S_{\mathbf{k}a\mathbf{k}'a'}|^2 = 
|t_{\mathbf{k}a\mathbf{k}'a'}|^2
  \frac{|(v_{\mathbf{k}'a'}^{\mathrm{out}})_z|}
{|(v_{\mathbf{k}a}^{\mathrm{in}})_z|} 
\lbl{eqTfitfi}
\end{equation}
This point will be addressed in detail in the following section.

%%%%%%%%%%%%%%%%%%%%%%%%%%%%%%%%%%%%%%%%%%%%%%%%%%%%%%%%%%%%%%%%%%%%%%
\section{Connection between Kubo and Landauer Approaches\lbl{SecKL}}

The connection of the Landauer approach involving the $S$-matrix to
the Kubo-Greenwood conductivity formula involving the Green functions
has been given by Fisher and Lee\cite{Fisher81}, Stone and
Szafer\cite{Stone88}, and Baranger and Stone\cite{Baranger89} in the
case of free electrons in the leads. However, for crystalline leads one
must account for the relevant band structure. Here we present some
proofs needed to extend the above results to crystalline leads. In
particular we shall pursue the $S$-matrix elements between in-states
from L and out-states in R and see how these connect to the Green
function of the system and to the conductance. In this way the
connection of the Kubo to the Landauer formalism will be made. We
proceed in three steps: {\it (i)} we find an expression for the
transmission amplitude $t_{\mathbf{k}a;\mathbf{k'}a'}$; {\it (ii)} we
express the asymptotic Green function in terms of the $S$-matrix; and
{\it (iii)} we express the $S$-matrix in terms of
$t_{\mathbf{k}a;\mathbf{k'}a'}$.

{\it (i)} We start with the calculation of current matrix elements
between Bloch states. In particular, let
$\Psi_{\mathbf{k}a}^{\mathrm{in}}$ and
$\Psi_{\mathbf{k}a}^{\mathrm{out}}$ be incoming and outgoing Bloch
states (in the sense described before, {\it i.e.}~right- and
left-traveling) of the same energy
($E_{\mathbf{k}a}=E{\mathbf{k'}a'}$) in the same lead; the states are
normalized to unit probability rather than unit flux. Also, let $S$ be
the lead cross section, normal to the $z$-direction. Then the
following orthogonality relations hold for the current matrix
elements:
\begin{eqnarray}
\int_S dS {\Psi_{\mathbf{k} a}^{\mathrm{in}}}^*
\overleftrightarrow{\nabla}_z \Psi_{\mathbf{k}'a'}^{\mathrm{in}} &=&
\int_S dS {\Psi_{\mathbf{k} a}^{\mathrm{out}}}^*
\overleftrightarrow{\nabla}_z \Psi_{\mathbf{k}'a'}^{\mathrm{out}} \nonumber\\
&=& i\frac{2m}{\hbar} |(v_{\mathbf{k} a})_z| 
\delta_{\mathbf{k}\mathbf{k}'}\delta_{aa'} 
\lbl{eqOrtha} \\
\int_S dS {\Psi_{\mathbf{k} a}^{\mathrm{in}}}^*
\overleftrightarrow{\nabla}_z \Psi_{\mathbf{k}'a'}^{\mathrm{out}} &=& 0
\lbl{eqOrthb} \\
\mbox{(for $E_{\mathbf{k}'a'}=E_{\mathbf{k}a}$)} \nonumber
\end{eqnarray}
(${\Psi_{\mathbf{k} a}^{\mathrm{in}}}$ and ${\Psi_{\mathbf{k}'
    a'}^{\mathrm{out}}}$ here are supposed to be left- and
right-traveling states in the same lead).  These have been proven
before\cite{Baranger89} for free electrons; we present the proof for
Bloch states in Appendix~\ref{appA}.

Eqs.~(\ref{eqOrtha},\ref{eqOrthb}) can be used to project a scattering
wavefunction onto a particular channel; using them in connection with
Eq.~(\ref{eqScattering}) we can extract the transmission amplitude out
of $\Psi_{\,\mathbf{k}a}^{\mathrm{tot}}$:
\begin{equation}
t_{\mathbf{k}a;\mathbf{k'}a'}\cdot i\frac{2m}{\hbar}
|(v_{\mathbf{k'}a'}^{\mathrm{out}})_z|  = 
\int_S dS \Psi_{\,\mathbf{k}a}^{\mathrm{tot}*}(\mathbf{r}) 
\overleftrightarrow{\nabla}_z
\Psi_{\mathbf{k}'a'}^{\mathrm{R\,out}}(\mathbf{r})
\lbl{eqtampl}
\end{equation}

{\it (ii)} Next we consider the asymptotic expression for the retarded Green
function $G^+(\mathbf{r},\mathbf{r}';E)$, with $\mathbf{r}$ and
$\mathbf{r}'$ in different leads and going to infinity. This can be
found\cite{Baranger89} by using the boundary condition that
$G^+(\mathbf{r},\mathbf{r}';E)$ should represent an outgoing wave at
$\mathbf{r}$, $G^-(\mathbf{r},\mathbf{r}';E)$ should represent an
incoming wave at $\mathbf{r}$, and 
\begin{equation}
G^+(\mathbf{r},\mathbf{r}';E) =G^{-*}(\mathbf{r}',\mathbf{r};E).
\label{eqGretGadv}
\end{equation}
The Green function then, expanded in in- and out-states, has the
asymptotic form
\begin{eqnarray}
G^+(\mathbf{r},\mathbf{r}';E) = \sum_{\mathbf{k}a} \sum_{\mathbf{k}'a'}
A_{\mathbf{k}a;\mathbf{k}'a'} 
\Psi_{\mathbf{k}a}^{\mathrm{R\, out}}(\mathbf{r})
\Psi_{\mathbf{k}'a'}^{\mathrm{L\, in}*}(\mathbf{r}') \\
\mbox{(for $z\rightarrow \infty$, $z'\rightarrow -\infty$)} \nonumber
\lbl{eqAsymptgreen1}
\end{eqnarray}
We are now seeking a relation between the coefficients
$A_{\mathbf{k}a;\mathbf{k}'a'}$ and the transmission amplitudes
$t_{\mathbf{k}a;\mathbf{k}'a'}$. Eq.~(\ref{eqtampl}) was used together
with the Lippmann-Schwinger equation and Eq.~(\ref{eqBaranger}) in
Ref.~\onlinecite{Baranger89} to prove this for free-electron leads.
However, the Lippmann-Schwinger equation is difficult to handle when
we have different materials in the leads, because it connects the
Bloch wavefunctions of different materials in the infinite leads, thus
there is no localized perturbation. Alternatively, one can start
directly from the definition\cite{Newton82} of the $S$-matrix in terms
of the time-dependent retarded Green function (propagator), looking at
the transmission amplitude from an initial wavepacket
$\Phi^{\mathrm{L\, in}}_i(\mathbf{r},t)=\int dE_i e^{-iE_i t} a_i(E_i)
\Psi^{\mathrm{L\, in}}_{i}(\mathbf{r};E_i)$, incident from the left,
to a final one $\Phi^{\mathrm{R\, out}}_f(\mathbf{r},t)=\int dE_f
e^{-iE_f t} a_f(E_f) \Psi^{\mathrm{R\, out}}_{f}(\mathbf{r};E_f)$,
outgoing to the right.  Here $\Psi_i$ and $\Psi_f$ denote itinerant
Bloch states and we have propagated the initial and final states to
$t$ and $t'$ by their corresponding bulk hamiltonians giving the
exponential factors. At the end $a_i$ and $a_f$ will be taken as
extremely peaked distributions around the same energy $E_i$; $i$ and
$f$ represent then definite $\mathbf{k}'a'$ and $\mathbf{k}a$. One
gets
\begin{widetext}
\begin{eqnarray}
S_{fi} &:=&
\lim_{{t\rightarrow \infty}\atop{t'\rightarrow -\infty}}
\int d^3r \int d^3r' \Phi^{\mathrm{R\, out}*}_f(\mathbf{r},t)
G^+(\mathbf{r},t;\mathbf{r}',t') \Phi^{\mathrm{L\, in}}_i(\mathbf{r}',t') 
\nonumber\\
&=&
\lim_{{t\rightarrow \infty}\atop {t'\rightarrow -\infty}}
\int d^3r \int d^3r' \Phi^{\mathrm{R\, out}*}_f(\mathbf{r},t)
\int dE e^{-iE(t-t')} G^+(\mathbf{r},\mathbf{r}';E)
\Phi^{\mathrm{L\, in}}_i(\mathbf{r}',t') \nonumber\\
%&=&
%\lim_{{t\rightarrow \infty}\atop {t'\rightarrow -\infty}}
%\int d^3r \int d^3r' \Phi^{\mathrm{R\, out}*}_f(\mathbf{r},t)
%\int dE e^{-iE(t-t')} 
%\sum_{f'i'} A_{f'i'}(E) \Psi^{\mathrm{R\, out}}_{f'}(\mathbf{r};E)
%\Psi^{\mathrm{L\, in}*}_{i'}(\mathbf{r}';E)  
%\Phi^{\mathrm{L\, in}}_i(\mathbf{r}',t') \nonumber\\
&=&
\lim_{{t\rightarrow \infty}\atop {t'\rightarrow -\infty}}
\int dE e^{-iE(t-t')} \sum_{f'i'} A_{f'i'}(E) \nonumber\\
&&\phantom{\lim_{{t\rightarrow \infty}\atop {t'\rightarrow -\infty}}}
\int dE_f e^{iE_f t} a_f^*(E_f) 
(\Psi^{\mathrm{R\, out}}_f(E_f),\Psi^{\mathrm{R\, out}}_{f'}(E))
\int dE_i e^{-iE_i t'} a_i(E_i) 
(\Psi^{\mathrm{L\, in}}_{i'}(E),\Psi^{\mathrm{L\, in}}_i(E_i))\nonumber\\
%&=&
%\lim_{{t\rightarrow \infty}\atop {t'\rightarrow -\infty}}
%\int dE e^{-iE(t-t')} \sum_{f'i'} A_{f'i'}(E) \int dE_i dE_f a_f^*(E_f) a_i(E_i)
%e^{iE_f t} \delta_{ff'} v_f \delta(E-E_f)
%e^{-iE_i t'}\delta_{ii'} v_i \delta(E-E_i)\nonumber\\
&=&
\int dE a_f^*(E) a_i(E) v_i v_f A_{fi}(E) \label{eqSfi_a}\\
&=&
A_{fi}(E_i) \sqrt{v_i v_f} \ \ \ \ \ \ \mbox{(on-shell)}.
\lbl{eqSfi}
\end{eqnarray}
\end{widetext}
The velocities appearing represent the $z$ component of the group
velocity, since in waveguide geometry the $\mathbf{k}_{\parallel}$'s
form in reality a very dense discreet set so that the wavepacket is
constructed for a definite $\mathbf{k}_{\parallel}$ from the
continuous $k_z$ spectrum.  Having this in mind, we have used in the
derivation of (\ref{eqSfi_a}) the orthonormality condition for Bloch
waves
\begin{eqnarray}
  (\Psi_{\mathbf{k}a}(E_f),\Psi_{\mathbf{k}'a'}(E))
&:=&
\int d^3r \Psi_{\mathbf{k}a}^*(\mathbf{r},E_f)\Psi_{\mathbf{k}'a'}(\mathbf{r};E) 
\nonumber\\
&=&
  \delta_{\mathbf{k}_{\parallel}\mathbf{k}_{\parallel}'}\delta_{aa'}
 \delta(k_z-k'_z)\nonumber\\ 
&=&
  \delta_{\mathbf{k}_{\parallel}\mathbf{k}'_{\parallel}} 
  \delta_{aa'}v_z \delta(E-E_f).
\lbl{eqNorm1}
\end{eqnarray}
In addition we have used the normalization of wavepackets to unit
probability
\begin{eqnarray}
1 &=& \int dE a^*(E) \int dE' a(E') (\Psi(E),\Psi(E')) \nonumber\\
&=& \int dE |a(E)|^2 v_E
\lbl{eqNorm2}
\end{eqnarray}
which for a very peaked distribution $a(E)$ around some energy $E_i$ gives
\begin{equation}
\int dE |a(E)|^2 = 1/v_{E_i}.
\lbl{eqNorm3}
\end{equation}
implying that in the limiting case $|a(E)| \rightarrow \sqrt{\delta(E-E_i)
  /v_{E_i}}$. Finally, in the last step we have assumed that $a_f(E)$
and $a_i(E)$ are both peaked around the same energy $E_i$ so that the
wavepacket goes to a single Bloch function, whence $a_f^*(E)
a_i(E)\rightarrow \delta(E-E_i) / \sqrt{v_f(E_i) v_i(E_i)}$. Thus, the
final result (\ref{eqSfi}) is valid for on-energy-shell scattering of
Bloch waves; else, for general wavepackets, Eq.~(\ref{eqSfi_a}) must
be applied. Working with wavepackets has guaranteed the correct
normalization.

From Eq.~(\ref{eqSfi}) we see that the coefficients in the Green
function asymptotic expansion are just $S$-matrix elements (normalized
to the group velocities). In this form, the $S$-matrix is unitary, {\it
i.e.}~the scattering probability is
\begin{equation}
T_{fi}=|S_{fi}|^2.
\lbl{eqTfi.Sfi}
\end{equation}

{\it (iii)} Finally we show that the relation of the $S$-matrix to the
previously defined transmission amplitude of Eq.~(\ref{eqScattering})
is
\begin{equation}
S_{fi} = t_{fi}\frac{\sqrt{v_f}}{\sqrt{v_i}}=
A_{fi} \sqrt{v_f v_i}
\lbl{eqSfi.tfi}
\end{equation}
{\it i.e.}, a normalization involving the group velocities is needed.
This can be seen by noting that an incoming wavepacket
$\Phi_{\mathrm{in}} = \int dE a(E) \Psi_{\mathrm{in}}(E)$, normalized
to unit probability as in Eq.~(\ref{eqNorm2}), evolves partly into the
wavepacket $\Phi_{\mathrm{out}} = \int dE \, t(E)
a(E)\Psi_{\mathrm{out}}(E)$ according to Eq.~(\ref{eqScattering}).
Assuming that $a(E)$ is so much peaked around $E_0$ that $t$ and $v$
are constant in this energy range, the scattering probability is given
by the normalization factor of the outgoing wavepacket:
$T_{fi}=||\Phi_{\mathrm{out}}||^2 = |t|^2 \int dE \int dE' |a(E)|^2
(\Psi_{\mathrm{out}}(E), \Psi_{\mathrm{out}}(E')) = |t|^2 \int dE
|a(E)|^2 \int dE' \delta(E-E')v_{\mathrm{out}} = |t|^2
v_{\mathrm{out}}/v_{\mathrm{in}}$, where the normalizations
(\ref{eqNorm1}) and (\ref{eqNorm3}) have been utilized. Comparing this
to Eq.~(\ref{eqTfi.Sfi}) proves Eq.~(\ref{eqSfi.tfi}) up to a phase
factor, which can be seen to be just unity by creating a wavepacket
out of $\Psi^{\mathrm{tot}}$ in Eq.~(\ref{eqScattering}) and
constructing the $S$-matrix element. Furthermore the expression
(\ref{eqTfitfi}) results from (\ref{eqTfi.Sfi}) and (\ref{eqSfi.tfi}).

Combining all the above, we may rewrite Eq.~(\ref{eqAsymptgreen1}) as:
\begin{eqnarray}
G^+(\mathbf{r},\mathbf{r}';E) 
&=& \sum_{\mathbf{k}a} \sum_{\mathbf{k}'a'}
\frac{S_{\mathbf{k}a;\mathbf{k}'a'}}{\sqrt{(v_{\mathbf{k}'a'})_z(v_{\mathbf{k}a}})_z}
\Psi_{\mathbf{k}'a'}^{\mathrm{L\, in}*}(\mathbf{r}')
\Psi_{\mathbf{k}a}^{\mathrm{R\, out}}(\mathbf{r}) \nonumber \\
&=& \sum_{\mathbf{k}a} \sum_{\mathbf{k}'a'}
\frac{t_{\mathbf{k}a;\mathbf{k}'a'}}{(v_{\mathbf{k}'a'})_z}
\Psi_{\mathbf{k}'a'}^{\mathrm{L\, in}*}(\mathbf{r}')
\Psi_{\mathbf{k}a}^{\mathrm{R\, out}}(\mathbf{r}) \lbl{eqAsymptgreen2}
\end{eqnarray}
which gives us for the advanced Green function
\begin{equation}
G^-(\mathbf{r},\mathbf{r}';E) = \sum_{\mathbf{k}a} \sum_{\mathbf{k}'a'}
\frac{t_{\mathbf{k}a;\mathbf{k}'a'}^*}{(v_{\mathbf{k}'a'})_z}
\Psi_{\mathbf{k}'a'}^{\mathrm{R\, out}*}(\mathbf{r}')
\Psi_{\mathbf{k}a}^{\mathrm{L\, in}}(\mathbf{r}).
\lbl{eqAsymptgreen3}
\end{equation}
We can extract the $S$-matrix elements from the Green function:
\begin{eqnarray}
&&S_{\mathbf{k}a;\mathbf{k}'a'} = \frac{\hbar^2}{4m^2}\,
\frac{1}{\sqrt{(v_{\mathbf{k}'a'})_z(v_{\mathbf{k}a}})_z} \lbl{eqSgreen}\\
&&
\times\int_{S_L} dS' \int_{S_R} dS
\Psi^{\mathrm{L\, in}}_{\mathbf{k}'a'}(\mathbf{r}') 
\overleftrightarrow{\nabla}'_{z'}
G^+(\mathbf{r},\mathbf{r}';E)
\overleftrightarrow{\nabla}_z
\Psi^{\mathrm{R\, out}*}_{\mathbf{k}a}(\mathbf{r}) 
\nonumber
\end{eqnarray}
where $S_L$ and $S_R$ have to be sufficiently far apart so that the
asymptotic formula (\ref{eqAsymptgreen1}) without evanescent states
can be used (the evanescent states are always included in the
self-consistent Green function). 

As a final result, we combine the above steps {\it (i)-(iii)} to
arrive at the Landauer formula. We start from the Baranger-Stone
expression~(\ref{eqBaranger}), expand the Green functions according to
(\ref{eqAsymptgreen2}) and (\ref{eqAsymptgreen3}), and use the
orthogonality relations (\ref{eqOrtha}) to get rid of some terms:
\begin{eqnarray}
g &=&  - \frac{e^2 \hbar^3}{16\pi m^2} \int_{S_{\mathrm{L}}} dS\int_{S_{\mathrm{R}}} dS'
\nonumber\\
&&\times
G^+(\mathbf{r},\mathbf{r}';E_F) 
\overleftrightarrow{\nabla}_z \overleftrightarrow{\nabla}'_z
G^-(\mathbf{r}',\mathbf{r};E_F) \nonumber\\
&=& \frac{e^2}{h}
\sum_{\mathbf{k} a\,\mathrm{in}}\sum_{\mathbf{k}' a'\,\mathrm{out}}
|t_{\mathbf{k} a;\mathbf{k}' a'}|^2
\frac{|(v_{\mathbf{k}' a'})_z|}{|(v_{\mathbf{k} a})_z|} \nonumber\\
&=& \frac{e^2}{h}
\sum_{\mathbf{k} a\,\mathrm{in}}\sum_{\mathbf{k}' a'\,\mathrm{out}}
T_{\mathbf{k}' a';\mathbf{k} a}
\end{eqnarray}

%%%%%%%%%%%%%%%%%%%%%%%%%%%%%%%%%%%%%%%%%%%%%%%%%%%%%%%%%%%%%%%%%%%%%%
\section{The conductance formula in KKR \lbl{SecKKR}}

\subsection{The Green function in the KKR formalism}

In the KKR method the one-electron retarded Green function is expanded
in terms of local orbitals centered at the atomic sites
$\mathbf{R}_n$ as:
\begin{eqnarray}
&&G^+(\mathbf{R}_n+\mathbf{r},\mathbf{R}_{n'}+\mathbf{r}';E) \nonumber\\
&&= -i\sqrt{E}\sum_L R_L^n(\mathbf{r}_{<};E)
H_L^{n'}(\mathbf{r}_{>};E) \delta_{nn'}
\nonumber\\
&&+ \sum_{LL'} R_L^n(\mathbf{r};E) G_{LL'}^{nn'}(E)R_L^{n'}(\mathbf{r}';E)
\lbl{eqKKR1}
\end{eqnarray}
Here, $R_L^n(\mathbf{r};E)$ and $H_L^{n'}(\mathbf{r};E)$ are,
respectively, the regular and irregular solutions of the Schr\"odinger
equation for the single potential $V_n(\mathbf{r})$ of the $n$th cell
in free space. Atomic units are used ($e=-\sqrt{2}$, $\hbar=1$,
$m=1/2$). The index $L=(l,m)$ represents the angular momentum quantum
number. The position vector $\mathbf{r}$ is confined in the atomic
cell $n$; $\mathbf{r}_{>}$ and $\mathbf{r}_{<}$ are the longer and
shorter, respectively, of $\mathbf{r}$ and $\mathbf{r}'$. In
Eq.~(\ref{eqKKR1}) the first term gives the on-site contribution to
the Green function, while the second is the so-called backscattering
term, where the information on the intersite electron propagation is
contained in the structure constants $G_{LL'}^{nn'}(E)$. These are
related to the known structure constants of a reference system, {\it
  e.g.}~vacuum, via an algebraic Dyson equation which includes the
local $t$-matrix of each single-sited potential. For further details
we refer to Refs.~\onlinecite{Wildberger97}.

The layered systems that shall be considered with the KKR formalism
consist of two half-infinite crystalline leads, assumed to have
perfect periodicity otherwise. Sandwiched between these leads is an
``interaction'' region where a different material can be placed and
where the scattering of the Bloch waves takes place. Three cases can
be handled in this respect: {\it (i)} systems with in-plane
periodicity;\cite{Mavropoulos02} {\it (ii)} wire-like structures
embedded in vacuum (or in some other non-conducting
medium);\cite{Opitz02,Papanikolaou02b} and {\it (iii)} atomic
constrictions between semi-infinite leads.

Concerning case {\it (i)}, when we consider systems with
two-dimensional in-plane ($x$-$y$) periodicity (perpendicular to the
direction of growth $z$), the interaction region and the two leads
have common in-plane Bravais vectors. If needed, larger
(non-primitive) two-dimensional unit cells are taken to match the
lattice constants of both materials; this is the case, for example, in
an Fe/GaAs contact.\cite{Mavropoulos02} The two-dimensional
periodicity of the layered systems allows to Fourier-transform the
Green function in the $x$ and $y$ directions, obtaining a
two-dimensional Bloch vector $\mathbf{k}_{\parallel}=(k_x,k_y)$ as a
good quantum number, and retaining an index $i$ to characterize the
layer in the direction of growth $z$. The Green function connecting
the layers $i$ in the left lead and $i'$ in the right lead is then
written
\begin{eqnarray}
\lefteqn{
G^+(\mathbf{R}_i+\chi_{\nu}+\mathbf{r},\mathbf{R}_{i'}+\chi_{\nu'}+\mathbf{r}';E)}&&
\nonumber\\
&=&\frac{1}{S_{\mathrm{SBZ}}}
\int_{\mathrm{SBZ}} d^2k_{\parallel} \,
e^{i\mathbf{k}_{\parallel}(\chi_{\nu}-\chi_{\nu'})} \nonumber\\
&&\times\sum_{LL'} R_L^i(\mathbf{r};E)
G_{LL'}^{ii'}(\mathbf{k}_{\parallel};E) R_{L'}^{i'}(\mathbf{r}';E)
\lbl{eqKKR2}
\end{eqnarray}
where $\mathbf{\chi_{\nu}}$ and $\chi_{\nu'}$ are in-plane lattice
vectors, $\mathbf{R}_i$ is the interlayer lattice vector, SBZ is the
surface Brillouin zone of the system and $S_{\mathrm{SBZ}}$ its area.
In this equation each layer $i$ is assumed to have a unique atom type,
hence only the index $i$ suffices to characterize the local
wavefunction. In the case of more inequevalent atoms per layer, an
extra index $\mu$ can be introduced to account for the propagation
between different kinds of atoms. In the case of spin magnetism, the
Green function is different for each spin direction $\sigma=\uparrow$
or $\downarrow$. The formalism can be generalized for fully
relativistic calculations, where the spin-orbit coupling results in a
mixing of the two spin channels.

Once Eq.~(\ref{eqKKR1}) or Eq.~(\ref{eqKKR2}) is substituted into the
expression (\ref{eqBaranger}), with $\mathbf{r}$ and $\mathbf{r}'$ in
different leads, the on-site term of the Green function does not
contribute, and only the intersite term survives; moreover, the
spacial derivative affects only the local orbital functions
$R_L^n(\mathbf{r})$ and $H_L^{n'}(\mathbf{r})$, leaving the structure
constants untouched. Finally, if Eq.~(\ref{eqKKR2}) is used, the
conductance appears at first $\mathbf{k}_{\parallel}$-resolved,
$g(\mathbf{k}_{\parallel})$, which is most convenient in structures
with two-dimensional periodicity. A
$\mathbf{k}_{\parallel}$-integration then gives the result
\begin{equation}
g = \frac{1}{S_{\mathrm{SBZ}}}\int_{\mathrm{SBZ}} d^2k_{\parallel} \,
g(\mathbf{k}_{\parallel})
\end{equation}

In Eq.~(\ref{eqBaranger}), both the retarded and the advanced Green
functions are needed; however, they are related through the identity
(\ref{eqGretGadv}) for real $E$. This is used in our formulation; the
energy $E$ is identified in the calculations with the Fermi level
$E_F$ plus an (in principle infinitesimal) imaginary part $\epsilon$
which we take very small.

The case {\it (ii)} of wire-like structures is completely analogous,
but the Fourier transform of Eq.~(\ref{eqKKR2}) is not performed. More
specifically, consider a wire embedded in vacuum (see also
Fig.~\ref{figWire}); the vacuum region will be also divided
artificially in volume-filling cells. One can also consider defects
within the wire, or even two wires attached to some cluster of atoms
(as indicated in Fig.~\ref{figWire}), and solve self-consistently for
the electronic structure.  Due to electronic states at the surface of
the wire, the first one or two vacuum layers above the wire surface
can contribute to the conductance, but after that the cross section
can be truncated.  The cross sections left and right, where the
conductivity tensor must be evaluated, consist always of more than one
atomic cells, since one must include the vacuum region. In this way
the Green function must be considered for all combinations between
cells on the left and the right, and the expression~(\ref{eqBaranger})
for the conductance splits up in partial contributions corresponding
to these combinations:
\begin{equation}
g = \sum_{\mu'\mathrm{(Left)}} \sum_{\mu\mathrm{(Right)}} g_{\mu\mu'}
\label{eq:37}
\end{equation}
with $g_{\mu\mu'}$ given by Eq.~(\ref{eqBaranger}) but integrated over
single atomic cells. 

Analogoulsy, in case {\it (iii)} a similar setup can be used if we
consider semi-infinite 2D leads but current flow localized in space,
as in the case of transport through a constriction. In this case we
consider two planes as shown in Figure~\ref{figWire} (right), while
convergence must be checked with respect to the size of the regions
considered in the summations of Eq.~\ref{eq:37}

%In case that the two ends of the wire have different orientation, so
%that the direction of growth and propagation is, say, $\hat{z}$ for
%the region L but $\hat{z}'$ for the region R, the formalism still
%allows the calculation of the conductance through such a ``corner'',
%even though in the self-consistent calculation the direction $\hat{z}$
%is the unique $z$-direction throughout the system. In this case, in
%the region R one would have to compute the derivative with respect to
%$z'$ for the current matrix elements $J_{LL'}$. One has merely to know
%the rotation matrix $\mathbf{D}$ which transforms $\hat{z}'$ to
%$\hat{z}$, and calculate its representation matrix in angular momentum
%space as $D_{LL'} = \int d\Omega\, Y_L(\hat{r})
%Y_{L'}(\mathbf{D}\cdot\hat{r})$. This matrix $D_{LL'}$ can then be
%applied to the orbitals $R_L$ of the second lead, which are originally
%expressed in the initial coordinate system ($x$,$y$,$z$) to transform
%them in the local coordinate system ($x'$,$y'$,$z'$) and calculate
%$J_{LL'}$.

\begin{figure}[t]
\begin{center}
\includegraphics[angle=270,width=8cm]{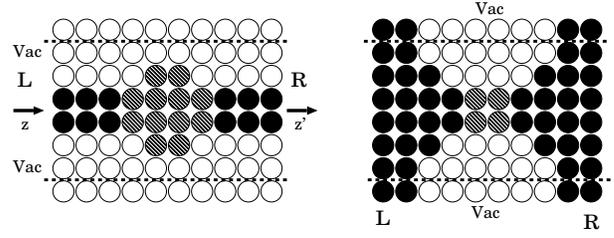}
\end{center}
\caption{Left: Setup for the calculation of conductance in a nanowire
configuration. The filled circles represent the wire atoms (leads),
including defects or an atomic cluster (shaded circles). The white
circles represent the vacuum region. The conductance calculation is
truncated at the dashed lines, assuming that the states of the
nanowire have decayed beyond this region. The arrows labeled $z$ and
$z'$ represent the directions of growth of the two leads. Right:
Similar setup for the calculation in the case of nano-size
constrictions between infinite leads. Again a truncation at the dashed
lines is taken, assuming that outside this region there is
insignificant tunneling between the leads.} \lbl{figWire}
\end{figure}

In the next subsections we will consider the calculation of the
spacial derivative of the local radial functions and the Green
function.

\subsection{Plane integration}

Firstly we consider a direct evaluation of the conductance by use of
Eq.~(\ref{eqBaranger}) and calculation of the spacial derivative of
the Green function at exactly the plane surfaces left
($S_{\mathrm{L}}$) and right ($S_{\mathrm{R}}$). Both $S_{\mathrm{L}}$
and $S_{\mathrm{R}}$ are assumed to be in the asymptotic region where
the potential is stabilized to the bulk one and the evanescent states
have decayed; in practice one has to perform the calculations for
several positions of $S_{\mathrm{L}}$ and $S_{\mathrm{R}}$ at finite
distances to verify that the results remain unchanged.

The set of atomic cells is volume-filling, and the plane surface
cutting through them inherits the cellular structure used in the KKR
method; thus it is split in two-dimensional tesselating cells. Each
one of them is a convex polygon corresponding to the section of the
plane that belongs to a convex Voronoi polyhedron\cite{footnote2} (or
just Wigner-Seitz cell in the monoatomic case). In this way, a
two-dimensional cellular Voronoi construction is defined in the plane,
each cell of which is completely within some three-dimensional cell.
Therefore the representation of the Green function in terms of local
radial functions can be readily used. In fact, in systems with
two-dimensional periodicity, a two-dimensional unit cell consisting of
some convex Voronoi polygons $S_{\mu}$ is constructed and the calculations
can be confined in those. An example of the construction for bcc (001)
surface cells is given in Fig.~(\ref{figVoronoi}).

\begin{figure}
\begin{center}
\resizebox{5cm}{!}{\includegraphics{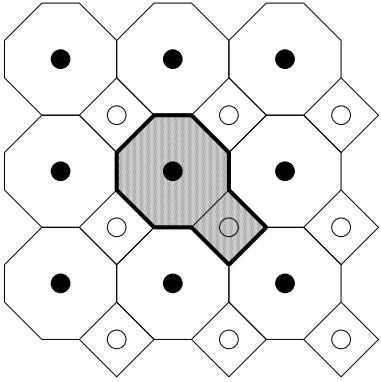}}
\end{center}
\caption{Two-dimensional geometrical construction for the bcc (001)
  surface cut. The cutting plane goes through the lattice sites of an
  atomic layer (full circles), but must include also part of the
  Wigner-Seitz cell of the next layer (open circles). The shaded area
  shows the two-dimensional unit cell formed, consisting of two
  smaller convex Voronoi polygons.} \lbl{figVoronoi}
\end{figure}

Using such a construction, the $\mathbf{k}_{\parallel}$-resolved
conductance is written as:
%\begin{widetext}
\begin{eqnarray}
g(\mathbf{k}_{\parallel}) =  
-\frac{1}{4\pi^3}
\sum_{\mu\mu'}
\sum_{LL'}\sum_{L''L'''} 
\left( J_{LL''}^{\mu} - J_{L''L}^{\mu*}
\right)\cdot \nonumber\\
\left(  
 J_{L'L'''}^{\mu'}- J_{L'''L'}^{\mu'*}
\right)
G_{LL'}^{\mu\mu'}(\mathbf{k}_{\parallel};E) 
G_{L''L'''}^{\mu\mu'*}(\mathbf{k}_{\parallel};E)
\lbl{eqBaranger3}
\end{eqnarray}
%\end{widetext}
where we have introduced the KKR current matrix elements
$J_{LL'}^{\mu}$(in a similar fashion as has been done in the
past\cite{Butler85,Banhart94}) in a cell as:
\begin{equation}
J_{LL'}^{\mu} = \int_{S_\mu} d^2r\, R_L(\mathbf{r};E_F) \partial_z
R^*_{L'}(\mathbf{r};E_F).
\lbl{eqCurrentPlane1}
\end{equation}
The summation $\sum_{\mu\mu'}$ is over the Voronoi polygons of the
inequevalent atomic sites of the 2-D unit cells in the leads. The
calculation of $J_{LL'}^{\mu}$ is described in Appendix \ref{appB}. In
the case of a finite system, where a two-dimensional Fourier transform
is not necessary, the summation is over all atoms in the planes
$S_{\mathrm{L}}$ and $S_{\mathrm{R}}$, and the
$\mathbf{k}_{\parallel}$ dependence drops.

The Green function matrix elements $G_{LL'}^{\mu\mu'}$, containing the
information for the propagation from site $\mu'$ at a layer within the
left lead to a site $\mu'$ at a layer in the right lead, are
off-diagonal, while for the charge density one needs only the diagonal
(on-site) elements. Nevertheless, an efficient O($N$) algorithm
exists\cite{Godfrin91} for their calculation within the screened KKR
formalism, {\it i.e.}, the time needed for the calculation scales
linearly with the distance between the two layers. Thus it is possible
to calculate the conductance in junctions of more than 100 monolayers
with present-day computers.\cite{Wunnicke02,Mavropoulos02}

\subsection{Volume integration: ASA and full cell}

In this subsection we provide an alternative to the calculation of the
surface-integrated current matrix elements of
Eqs.~(\ref{eqCurrentPlane1}) and (\ref{eqCurrentPlane2}). We prove
that the calculation can involve a volume integration over the unit
cell, instead of a surface one; in principle the results must be
equivalent, but this method has advantages when one wishes to use the
atomic sphere approximation. Most important is, however, that the
$l$-convergence is much better (see Sec.~\ref{SecExamples}).

First we observe that the value of the conductance is indeed
independent of the position of the planes $S_{\mathrm{L}}$ and
$S_{\mathrm{R}}$. This can be proven using the fact that the
conductivity tensor is divergenceless (Eq.~\ref{eqDivergenceless}).
Say that $\mathbf{r}$ is on $S_{\mathrm{R}}$; if we consider a second
plane surface $S_{\mathrm{R}}'$ close to $S_{\mathrm{R}}$, we can
utilize Gauss's theorem in the volume $V$ enclosed by the two planes,
to convert a volume integral of Eq.~(\ref{eqDivergenceless}) in $V$
into a surface integral over $S_{\mathrm{R}}$, $S_{\mathrm{R}}'$, plus
side-areas. The construction is analogous to the one described in
Appendix~\ref{appA}, as shown in Fig.~\ref{figVol2surf}.  The
contribution from the side-areas vanishes because there we have either
totally confining boundary conditions or Born-von Karman boundary
conditions leading to cancelation from opposite side-areas due to the
opposite surface unit vector orientation; then we are left with
$\int_{S_{\mathrm{R}}} dS \sgm_{ij}(\mathbf{r},\mathbf{r}') \cdot
\hat{z} = \int_{S_{\mathrm{R}}'} dS \sgm_{ij}(\mathbf{r},\mathbf{r}')
\cdot \hat{z}$.  The same applies for $S_{\mathrm{L}}$, where $z'$
varies; thus the flux of $\sgm_{ij}(\mathbf{r},\mathbf{r}')$, {\it
  i.e.}~the conductance, is independent of the exact position of
$S_{\mathrm{L}}$ and $S_{\mathrm{R}}$, {\it q.e.d.}. In fact,
following these arguments, we see that $S_{\mathrm{L}}$ and
$S_{\mathrm{R}}$ do not even have to be planes; for instance, they can
follow the pattern of the Wigner-Seitz or Voronoi cells, as long as
they meet the requirement that they satisfy the Born-von Karman
periodic boundary condition in $x$ and $y$. In the case that they are
not planar surfaces, one must of course take the flux of the
conductivity tensor really along the normal $\hat{n}$ pointing
outward at each point of the surfaces, {\it i.e.}
\begin{equation}
g = \int_{S_{\mathrm{R}}} dS \int_{S_{\mathrm{L}}} dS' \hat{n}
\cdot\sgm(\mathbf{r},\mathbf{r'}) \cdot \hat{n}'
\lbl{eqConductance2}
\end{equation}

\begin{figure}
\begin{center}
\resizebox{5cm}{!}{\includegraphics{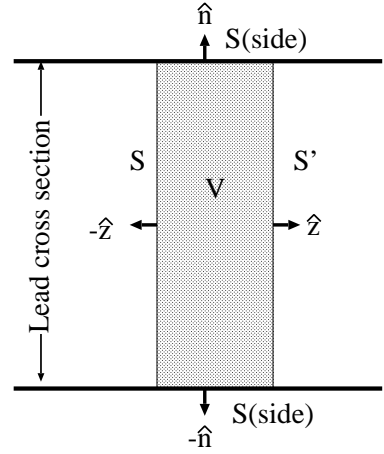}}
\end{center}
\caption{Construction for the conversion of volume to surface
  integrals over the lead cross section.}
\lbl{figVol2surf}
\end{figure}

Since the exact choice of $S_{\mathrm{R}}$ or $S_{\mathrm{L}}$ does
not affect the result, one can average over the volume $V$ included
between, say, $S_{\mathrm{R}}$ and $S_{\mathrm{R}}'$ instead of
integrating over $S_{\mathrm{R}}$. In particular, $V$ can be chosen to
have a thickness $d$ equivalent to a unit cell in $z$-direction, so
that one has to average over layer-adapted unit cells. In this
respect, the conductance formula has the same form as
Eq.~(\ref{eqBaranger3}), but with an extra prefactor of $1/d^2$ to
account for the volume averaging in the two leads ($d$ here is the
distance between two consecutive lattice monolayers); the current
matrix elements $J_{LL''}$, volume-averaged here in the atomic cells,
have the form of Eq.~(\ref{eqCurrentPlane1}) but with the integral
being three-dimensional over the atomic cell. This can be done both in
the atomic sphere approximation (ASA) and in the full-potential (and
full cell) formalism.  Details about their calculation are given again
in Appendix~\ref{appB}.

A word of caution is due here: it is essential that the volume
averaging leaves no ``holes'' in the structure. To be specific, assume
a bcc-like structure. Then, as is evident from Fig.~\ref{figVoronoi},
the Voronoi cells of the first layer (open circles) can touch via the
``holes'' (corresponding to the squares) to the cells of the third
layer. In this way, the current can partly by-pass one monolayer
traveling directly to the next one. If one just takes the current
averaged over the Voronoi cells or the ASA spheres of the middle
monolayer, one forgets to calculate this part of the current; this is
why the full many-atom unit cell must be taken, so that no such holes
are left.

If one uses the usual Wigner-Seitz cells (or atomic spheres),
Eq.~(\ref{eqCurrentVol1}) can induce a small inaccuracy. The reason is
that the volume constructed by such cells is in general not included
between planar surfaces, as in Fig.~(\ref{figVol2surf}), but rather
between corrugated surfaces, in accordance to the form of the
Wigner-Seitz cells. In such a case, the full conductivity tensor
$\sgm(\vc{r},\vc{r}')$ (not just the $\sgm_zz$ component) and
Eq.~(\ref{eqConductance2}) should be used in principle. To avoid such
a more complicated calculation, two ways can be followed, as
demonstrated in Fig.~\ref{figCellaverage}.
\begin{figure}
\begin{center}
\includegraphics[width=7cm]{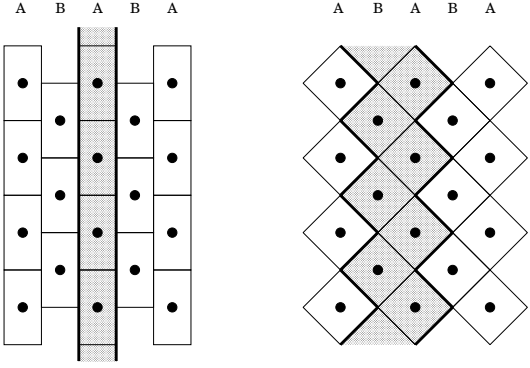}
\end{center}
\caption{Two possibilities for cell-averaging of the conductivity
  tensor flux in an ...ABAB... stacking sequence. Left, using a
  cellular division without corrugation of the surfaces, and right,
  using the Wigner-Seitz construction with corrugation. The latter is
  preferable in the KKR method.} \lbl{figCellaverage}
\end{figure}
Firstly, one can persist in using layer-adapted unit cells
(parellelepipeds), which give no corrugation.  This has the
disadvantage that such cells can be too flat so that the $l$ expansion
of the cell-centered KKR Green function and wavefunction converges
poorly.  Secondly, one can average over more than one monolayers; in
this example they would be A-A, A-B, B-A, and B-B. Then the corrugated
region is a smaller fraction of the total averaging volume, so that
the error becomes smaller. Test calculations on this will be given in
the Section~\ref{SecExamples} for bulk Al.

\subsection{Current matrix elements and selection rules}

Which method, volume or plane integration, is most convenient and
accurate depends on each specific problem; however one can have
``rules of thumb'' on the difficulty and convergence of each one. In
many systems the atomic sphere approximation is used, where the
potential around each site is assumed to be spherically symmetric, but
still a full multipole expansion of the charge density is taken. This
has the advantage of greater simplicity and less computational effort
than a full potential and full cell description. In such cases, the
plane integration is not applicable, since a plane cut through
volume-conserving spheres cannot give an accurate surface area; the
volume integration is then the only way. In the case of a full cell
treatment, when one has the correct Wigner-Seitz or Voronoi volume
tesselation, one must consider that the plane method has a drawback,
namely that the plane might go through regions only at the edge of
certain cells, where the $l_{\mathrm{max}}$ cutoff seriously affects
the accuracy of the results; on the other hand the volume integration
averages out such inaccuracies.

One has also ``selection rules'' that make certain $J_{LL'}$ elements
vanish. This is most easily seen if one uses spherical potentials. To
be specific, say that the plane goes through the atomic site at $z=0$;
then, in the plane integration one can easily see that the elements
$J_{LL'}$ are non-vanishing for $l'=1,3,5,\ldots$ when $l$ is even and
for $l'=2,4,6,...\ldots$ when $l$ is odd. On the other hand, in the
case of volume integration the $J_{LL'}$ are non-vanishing only for
$l'=l,l\pm 1$, {\it i.e.}~$J_{LL'}$ is band-diagonal in $l$ and
$l'$. This can be viewed as an advantage of the volume-averaging
method, since it means that, if one describes the electronic structure
with orbitals truncated at $l_{\mathrm{max}}$, for the accuracy of
$J_{LL'}$ one has to consider wavefunctions only up to
$l'=l_{\mathrm{max}}+1$. A full cell treatment adds more nonzero
elements, but the main contribution still comes from the ones
mentioned.

%%%%%%%%%%%%%%%%%%%%%%%%%%%%%%%%%%%%%%%%%%%%%%%%%%%%%%%%%%%%%%%%%%%%%%
\section{Examples \lbl{SecExamples}}

\subsection{Band counting in bulk conductance}

When the Landauer formula is applied to a perfectly periodic material,
{\it e.g.} the bulk of a crystal, it gives a finite conductance which
physically represents the conductance of a long wire placed between
two phase-randomizing electrodes.\cite{Imry99} Resolved in
$\mathbf{k}_{\parallel}$, the value of the conductance equals the
number of right-propagating (or equivalently left-propagating) states
at $E_F$ for this $\mathbf{k}_{\parallel}$. In other words, one has to
count the Fermi surface bands for that $\mathbf{k}_{\parallel}$, which
propagate in the direction $k_z \perp \mathbf{k}_{\parallel}$ with
$v_z > 0$. This is demonstrated in Fig.~\ref{figAlcond1}, where part
of the Fermi surface of Al is presented, in the $k_x$-$k_z$ plane,
together with the conductance in the $z$ direction as a function of
$k_x$ (with $k_y=0$). Actually, some of the bands shown have $v_z <
0$, but their equivalents with $v_z > 0$ exist symmetrically for
$k_z<0$. Clearly the conductance (in units of $e^2/h$) equals the
number of bands at $E_F$ for each $k_x$, giving a stepwise picture.

Also in Fig.~\ref{figAlcond1} we can compare the results for angular
momentum truncation at $l_{\mathrm{max}}=2$, $3$, and $4$. Increasing
$l_{\mathrm{max}}$ results in a more accurate description of the
wavefunction derivative.  Basically the slower convergence here arises
from the relatively high Fermi energy of Al, and is also present in a
free-electron approach for high $E_F$.  As noted in the previous
section, if the wavefunction is accurate for some $l_{\mathrm{max}}$,
for the derivative one has to take $l_{\mathrm{max}}+1$. For most
practical purposes, $l_{\mathrm{max}}=3$ is enough, considering also
that the calculation time for $l_{\mathrm{max}}=4$ is more than three
times the one of $l_{\mathrm{max}}=3$ (due to matrix inversion, the
calculation time scales roughly as $(l_{\mathrm{max}}+1)^6$).

In Fig.~\ref{figAlcond2} the conductance for the same system is shown,
but also analyzed in the various interlayer contributions. We see that
these can exhibit fluctuations in $\mathbf{k}_{\parallel}$, reflecting
fluctuations as a function of the interlayer distance. However, when
averaged over two monolayers on the left and two on the right, the
fluctuations practically cancel each other. The origin of these
fluctuations is the Wigner-Seitz construction for the unit cell,
resulting in corrugation of the surfaces where the current is
calculated. Due to the fact that we account only for the
$zz$-component of the conductivity tensor, when we have corrugation
the matrix elements in Eq.~(\ref{eqOrtha}) are not integrated
correctly and the nondiagonal current matrix element do not vanish;
thus, beating effects of the conductance appear. As the conductance is
averaged over more than one monolayers, the corrugation-free region in
the middle increases, the relative error due to the corrugation
decreases and the steps in the conductance become flat. If one would
use tetragonal or prismatic unit cells, suited for layered geometry,
then the fluctuations would be absent; however such cells are not
well-suited for an accurate calculation of the wavefunctions in KKR.
\begin{figure}
\begin{center}
\includegraphics[width=7cm]{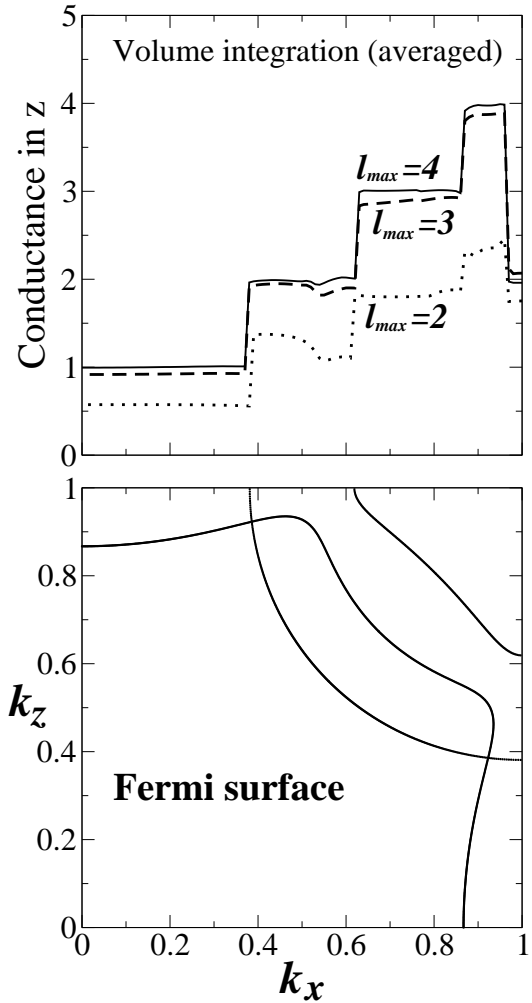}
\end{center}
\caption{Fermi surface in the $k_x$-$k_z$ plane (bottom) and
  conductance (in units of $e^2/h$) as a function of $k_x$ (top) for
  bulk Al and for ($l_{\mathrm{max}}=2$, $3$, and $4$).}
\lbl{figAlcond1}
\end{figure}
\begin{figure}
\begin{center}
\includegraphics[width=7cm]{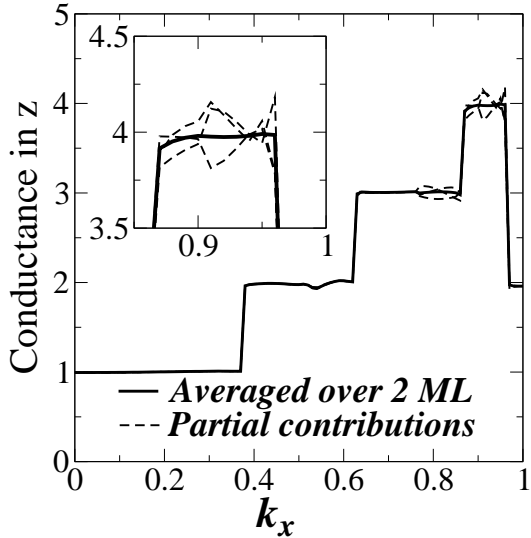}
\end{center}
\caption{Contributions to the fcc Al bulk conductance from the various
  layers (A-A, B-B, B-A, and A-B in the ...ABAB... stacking sequence
  of Fig.~\protect{\ref{figCellaverage}} right) (dashed lines) and
  averaged sum (full line), as a function of $k_x$, for
  $l_{\mathrm{max}}=4$.  The fluctuations in the partial contributions
  come from artificial beating effects due to corrugation, and vanish
  in the averaging procedure. The inset shows some of the fluctuations
  in more detail.} \lbl{figAlcond2}
\end{figure}

Finally, in Fig.~\ref{figAlcond3} we see a calculation for the same
system, but employing the in-plane integration, rather than the volume
averaging. Here the convergence with $l_{\mathrm{max}}$ is poor for
the reasons explained in the previous section; even for
$l_\mathrm{max}=4$ the deviations from integer conductance values are
large.
\begin{figure}
\begin{center}
\includegraphics[width=7cm]{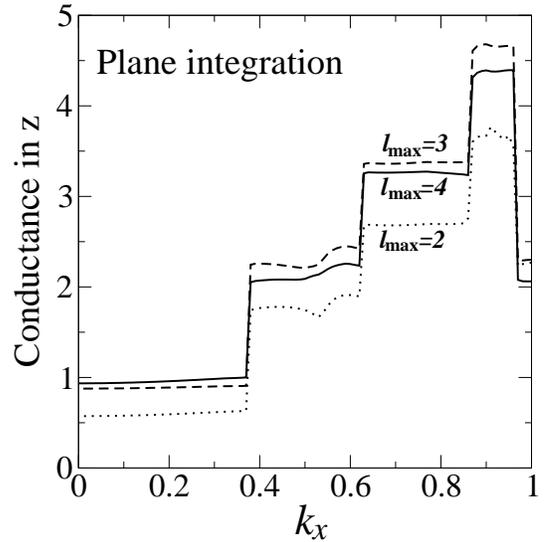}
\end{center}
\caption{
  Conductance as a function of $k_x$ (top) for bulk Al
  ($l_{\mathrm{max}}=2$, 3, and 4) in the case of plane-integration.
  It is not necessary to average over more monolayers, but the $l$
  convergence is rather poor and the integer valuer are not reached
  even for $l_{\mathrm{max}}=4$.} \lbl{figAlcond3}
\end{figure}

\subsection{The effect of the nonzero imaginary part of the energy \label{SecIme}}

Although the conductance should be calculated at a real energy $E$,
the Green function in the KKR method is always calculated at a complex
energy $E+i\epsilon$, and the real $E$ is approximated by taking very
small, but nonzero, $\epsilon$. This can have an artificial damping
effect to the conductance, since waves within a small energy range
around $E$ are effectively superimposed and finally the phase is
randomized, especially if the leads are seperated by a large distance.
In Fig.~\ref{figTemperature} we show an example of how small
$\epsilon$ should be in a realistic calculation. The system here
consists of two Fe leads with parallel magnetic moment seperated by a
ZnSe spacer. Electrons are injected from the first lead into the ZnSe
conduction band, and are detected by the second Fe lead. The spacer
thickness is varied from 9 to 97 monolayers, and due to the multiple
reflections at the two interfaces transmission resonances appear at
certain thicknesses. For more information we refer to
Ref.~\onlinecite{Mavropoulos02}. The conductance of the majority
electrons for $\mathbf{k}_{\parallel}=0$, {\it i.e.}, at the
$\bar{\Gamma}$-point, is presented in Fig.~\ref{figTemperature} for a
choice of $\epsilon=0.02$~mRy, 0.2~mRy, and 1~mRy. For
$\epsilon=1$~mRy there is strong artificial damping, while
$\epsilon=0.02$~mRy is adequate even for the large thickness of 97
monolayers. Note that this damping cannot model an effect of
temperature, because, when we depart from the real axis, the spectral
density of a state transforms from a delta function to a Lorenzian
distribution and not to the derivative of the Fermi function. Due to
the long tails of the Lorenzian the damping is much stronger than for
a Fermi distribution of the same halfwidth.

\begin{figure}
\begin{center}
\includegraphics[width=7cm]{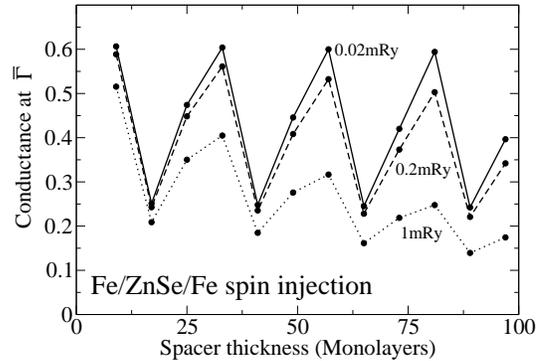}
\end{center}
\caption{The effect of the nonzero imaginary part $\epsilon$ of the
  energy. Conductance (in units of $e^2/h$) as a function of spacer
  thickness for spin injection through the conduction band in a
  Fe/ZnSe/Fe (001) junction. The oscillations are due to multiple
  reflections. For $\epsilon=1$~mRy there is strong artificial
  damping, while $\epsilon=0.02$~mRy is adequate even for the
  thickness of 97 monolayers.} \lbl{figTemperature}
\end{figure}

\section{Summary and Conclusions \label{SecSummary}}

We have presented a formalism for the calculation of ballistic
conductance in solids, based on the KKR Green function method for the
ground-state electronic structure combined with the
Landauer-B\"uttiker approach.  It makes use of the result of Baranger
and Stone\cite{Baranger89} connecting the derivative of the
one-electron Green function to the conductance. For the foundation of
the formalism, we have discussed the relation of the $S$-matrix
between Bloch in- and out-states to the conductance. We have given an
expression connecting the $S$-matrix to the Green function of the
system, generalizing the theory of Baranger and Stone to include the
realistic band structure of the leads.

The convergence of the method with angular momentum cutoff
($l_{\mathrm{max}}$) was studied and found to be comparable to that of
KKR.  It can be applied to systems with two-dimensional periodicity as
well as nanowires.  Our results show that the volume integration and
averaging of the current matrix elements, applicable both in ASA and
full-cell or full-potential approaches, gives well-converged results for
the calculation of ballistic transport. Owing to the linear scaling of
the calculational effort with the number of layers of the screened KKR
formalism (O($N$) scaling), our method is suitable for large systems.

\begin{acknowledgments}
  The authors are grateful to Professor N.~Stefanou for helpful and
  motivating discussions.  Moreover, financial support from the RT
  Network of {\em Computational Magnetoelectronics} (contract
  RTN1-1999-00145) of the European Commission is gratefully
  acknowledged.
\end{acknowledgments}

%%%%%%%%%%%%%%%%%%%%%%%%%%%%%%%%%%%%%%%%%%%%%%%%%%%%%%%%%%%%%%%%%%%%%%
\appendix

%%%%%%%%%%%%%%%%%%%%%%%%%%%%%%%%%%%%%%%%%%%%%%%%%%%%%%%%%%%%%%%%%%%%%%
\begin{widetext}
\section{Current matrix orthogonality relations \label{appA}}

Let $\Psi_{\mathbf{k} a}$ and $\Psi_{\mathbf{k'} a'}$ be two Bloch
wavefunctions of the same hamiltonian at the same energy. Then we
shall prove that the following relation holds:
\begin{equation}
\int_S dS (\mathbf{W}_{\mathbf{k}a;\mathbf{k}'a'})_z := 
\int_S dS (\Psi_{\mathbf{k} a}^* \overleftrightarrow{\nabla}_z
\Psi_{\mathbf{k}' a'}) =
i\frac{2m}{\hbar} |(v_{\mathbf{k}a})_z| 
\delta_{\mathbf{k}\mathbf{k}'}\delta_{aa'},
\lbl{eqA0}
\end{equation}
where $S$ is an (infinite) cross-sectional area in $x$ and $y$
directions.

The proof has as follows: First we note that, as a consequence of the
single-particle Schr\"odinger equation for a real potential,
\begin{equation}
\nabla \mathbf{W}_{\mathbf{k}a;\mathbf{k}'a'} = \nabla
(\Psi_{\mathbf{k} a}^* \overleftrightarrow{\nabla}
\Psi_{\mathbf{k}' a'}) 
= -\frac{2m}{\hbar} (E_{\mathbf{k}' a'} - E_{\mathbf{k} a})
\Psi_{\mathbf{k} a}^*\Psi_{\mathbf{k}' a'} = 0 \mbox{, for
$E_{\mathbf{k}' a'} = E_{\mathbf{k} a}$}
\lbl{eqA1}
\end{equation}
where the band indices $a$ and $b$ are used explicitly.  This is just
an expression for current conservation of Hamiltonian eigenstates.
Then, for each volume $V$ enclosed by a geometrical surface $S$,
Gauss's theorem gives
\begin{equation}
\oint_S dS \, \hat{n} \cdot \mathbf{W}_{\mathbf{k}a;\mathbf{k}'a'} = 
\int_V d^3r \, \nabla \mathbf{W}_{\mathbf{k}a;\mathbf{k}'a'} = 0
\lbl{eqA2}
\end{equation}
where $\hat{n}$ is a unit vector at the surface $S$ pointing outward.
In particular, $V$ can be chosen as a prismatic normal cross section
of the lead, extending from $z$ to $z+d$. Then its surface $S$ can be
decomposed in two plane cross sections $S_1$ at $z$ and $S_2$ at
$z+d$, as the bases of the prism, plus side-areas $S_{\mathrm{side}}$
at the lead surface, as shown in Fig.~\ref{figVol2surf}. At these side
areas we either have confining boundary conditions, {\it
  i.e.}~$\Psi_{\mathbf{k} a}|_{\mathrm{side}}=0$, whence
$\mathbf{W}_{\mathbf{k}a;\mathbf{k}'a'}|_{\mathrm{side}}=0$ or Born -
von Karman periodic boundary conditions, whence for each prism side
there is the opposite one with the \emph{same} value of
$\Psi_{\mathbf{k} a}|_{\mathrm{side}}$ and
$\mathbf{W}_{\mathbf{k}a;\mathbf{k}'a'}|_{\mathrm{side}}$ but with
\emph{opposite} orientation unit vector $\hat{n}$; then the sum of
their contributions to the surface integral will be again zero. In
this way, we are left with the two bases of the prism; they have
opposite unit vector orientations, thus:
\begin{equation}
\int_{S_1} dS_1 (-\hat{z})\cdot \mathbf{W}_{\mathbf{k}a;\mathbf{k}'a'} +
\int_{S_2} dS_2 \hat{z}\cdot \mathbf{W}_{\mathbf{k}a;\mathbf{k}'a'} = 0
\lbl{eqA3}
\end{equation}
or
\begin{equation}
\int_{S_1} dS_1 (\mathbf{W}_{\mathbf{k}a;\mathbf{k}'a'})_z =
\int_{S_2} dS_2 (\mathbf{W}_{\mathbf{k}a;\mathbf{k}'a'})_z
\lbl{eqA4}
\end{equation}
{\it i.e.}~$\int_{S} dS (\mathbf{W}_{\mathbf{k}a;\mathbf{k}'a'})_z$ is
independent of the position $z$ of the cross-section $S$. This means that
one can average it over the whole Born - von Karman supercell (in $z$)
of length $L$ and volume $V_{\mathrm{BvK}} = S\times L$:
\begin{equation}
\int_{S} dS (\mathbf{W}_{\mathbf{k}a;\mathbf{k}'a'})_z = 
\frac{1}{L} \int_0^L dz \int dS (\mathbf{W}_{\mathbf{k}a;\mathbf{k}'a'})_z(x,y,z) = 
\frac{1}{L} \int_{V_{\mathrm{BvK}}} d^3r 
(\mathbf{W}_{\mathbf{k}a;\mathbf{k}'a'})_z(x,y,z)
\lbl{eqA5}
\end{equation}
We now convert the volume integral in a sum over unit-cell integrals
$\sum_i\int_{V_0}d^3r$ and employ the Bloch property
\begin{equation}
\mathbf{W}_{\mathbf{k}a;\mathbf{k}'a'}(\mathbf{R}_i+\mathbf{r}) =
e^{i(\mathbf{k}'-\mathbf{k})\mathbf{R}_i}
\mathbf{W}_{\mathbf{k}a;\mathbf{k}'a'}(\mathbf{r})
\lbl{eqA6}
\end{equation}
to get:
\begin{eqnarray}
\int_{S} dS (\mathbf{W}_{\mathbf{k}a;\mathbf{k}'a'})_z(x,y,z) &=&
\frac{1}{L} \sum_i \int_{V_0}d^3r
(\mathbf{W}_{\mathbf{k}a;\mathbf{k}'a'})_z(\mathbf{R}_i+\mathbf{r}) \\
&=& \frac{N}{L} \delta_{\mathbf{k}\mathbf{k}'} \int_{V_0}d^3r
(\mathbf{W}_{\mathbf{k}a;\mathbf{k}'a'})_z(\mathbf{r})
\lbl{eqA7}
\end{eqnarray}
where $N$ is the total number of lattice sites in $V_{\mathrm{BvK}}$,
and
\begin{equation}
\sum_i e^{i(\mathbf{k}'-\mathbf{k})\mathbf{R}_i} = N
\delta_{\mathbf{k}\mathbf{k}'}
\lbl{eqA8}
\end{equation}
has been used; $V_0$ is the unit-cell volume. Note that $N V_0 =
V_{\mathrm{BvK}} = S L$, thus $N/L = S/V_0$, and we get:
\begin{equation}
\int_{S} dS (\mathbf{W}_{\mathbf{k}a;\mathbf{k}'a'})_z(x,y,z) = 
\frac{S}{V_0} \delta_{\mathbf{k}\mathbf{k}'} \int_{V_0}d^3r
(\mathbf{W}_{\mathbf{k}a;\mathbf{k}'a'})_z(\mathbf{r})
\lbl{eqA9}
\end{equation}
For $\mathbf{k}\neq\mathbf{k}'$ this gives zero, while for
$\mathbf{k}=\mathbf{k}'$ and $a=b$ we have
\begin{equation}
\int_{V_0}d^3r (\mathbf{W}_{\mathbf{k}a;\mathbf{k}a})_z(\mathbf{r}) =
\frac{im}{\hbar} \int_{V_0} d^3r 
(\Psi_{\mathbf{k} a}^* \frac{\hbar}{im} \nabla \Psi_{\mathbf{k} a} - 
 \Psi_{\mathbf{k} a} \frac{\hbar}{im} \nabla \Psi_{\mathbf{k} a}^*)
= i\frac{2m}{\hbar} \mathbf{v}_{\mathbf{k} a}
\lbl{eqA10}
\end{equation}
Eqs.~(\ref{eqA10}) and (\ref{eqA9}) verify (\ref{eqA0}) except in the
case of band crossing, when $\mathbf{k}=\mathbf{k}'$ but $a\neq a'$.
In this case we may use the identity connecting the momentum operator
$\mathbf{p}_{\mathrm{op}}$ to the Hamiltonian $H$ and the position
operator $\mathbf{r}_{\mathrm{op}}$:
\begin{equation}
\mathbf{p}_{\mathrm{op}} = \frac{im}{\hbar}
  [H,\mathbf{r}_{\mathrm{op}}]
\lbl{eqA11}
\end{equation}
For the evaluation of (\ref{eqA5}) we need the matrix element of
$\mathbf{p}_{\mathrm{op}}:=(\hbar/i)\nabla$. Using (\ref{eqA11}) we get
\begin{equation}
\frac{\hbar}{im}\int_{V_{\mathrm{BvK}}} d^3r \Psi_{\mathbf{k} a}^*
\mathbf{p}_{\mathrm{op}} \Psi_{\mathbf{k}' a'} =
(E_{\mathbf{k} a} - E_{\mathbf{k}' a'}) \int_{V_{\mathrm{BvK}}}
d^3r \Psi_{\mathbf{k} a}^* \mathbf{r} \Psi_{\mathbf{k}' a'}
\lbl{eqA12}  
\end{equation}
In case of band crossing, $(E_{\mathbf{k} a} - E_{\mathbf{k}' a'})=0$,
but this does not mean that the whole expression vanishes, since the
integral might diverge. It can be calculated in a standard way by
utilizing Bloch's theorem and reducing it to the unit cell. We have:
%\begin{eqnarray}
%\int_{V_{\mathrm{BvK}}}
%&d^3r& \Psi_{\mathbf{k} a}^*(\mathbf{r}) \mathbf{r} \Psi_{\mathbf{k}'
%  a'}(\mathbf{r}) \\
%&=& \sum_i \int_{V_0} d^3r 
%\Psi_{\mathbf{k} a}^*(\mathbf{R}_i+\mathbf{r})
%\,(\mathbf{R}_i+\mathbf{r})\, \Psi_{\mathbf{k}'
%  a'}(\mathbf{R}_i+\mathbf{r}) \\
%&=&\sum_i e^{i(\mathbf{k}'-\mathbf{k})\mathbf{R}_i}
%\int_{V_0} d^3r \Psi_{\mathbf{k} a}^*(\mathbf{r})
%\,(\mathbf{R}_i+\mathbf{r})\, \Psi_{\mathbf{k}' a'}(\mathbf{r}) \\
%&=&\sum_i e^{i(\mathbf{k}'-\mathbf{k})\mathbf{R}_i}
%\int_{V_0} d^3r \Psi_{\mathbf{k} a}^*(\mathbf{r})
%\,\mathbf{r}\, \Psi_{\mathbf{k}' a'}(\mathbf{r}) +
%\sum_i e^{i(\mathbf{k}'-\mathbf{k})\mathbf{R}_i} \mathbf{R}_i
%\int_{V_0} d^3r \Psi_{\mathbf{k} a}^*(\mathbf{r})
%\Psi_{\mathbf{k}' a'}(\mathbf{r}) \\
%&=&\frac{(2\pi)^3}{V_0} \delta (\mathbf{k}' - \mathbf{k})
%\int_{V_0} d^3r \Psi_{\mathbf{k} a}^*(\mathbf{r})
%\,\mathbf{r}\, \Psi_{\mathbf{k}' a'}(\mathbf{r}) +
%\frac{(2\pi)^3}{V_0} \nabla_{\mathbf{k}'} \delta (\mathbf{k}' -
%\mathbf{k})
%\int_{V_0} d^3r \Psi_{\mathbf{k} a}^*(\mathbf{r})
%\Psi_{\mathbf{k}' a'}(\mathbf{r})
%\lbl{eqA13}
%\end{eqnarray}
\begin{equation}
\int_{V_{\mathrm{BvK}}}
d^3r \Psi_{\mathbf{k} a}^*(\mathbf{r}) \mathbf{r} \Psi_{\mathbf{k}'
  a'}(\mathbf{r}) \\
= \sum_i \int_{V_0} d^3r 
\Psi_{\mathbf{k} a}^*(\mathbf{R}_i+\mathbf{r})
\,(\mathbf{R}_i+\mathbf{r})\, \Psi_{\mathbf{k}'
  a'}(\mathbf{R}_i+\mathbf{r}).
\end{equation}
Using the Bloch properties of $\Psi_{\mathbf{k} a}$ and $\Psi_{\mathbf{k}' a'}$ and the relations:
\begin{eqnarray}
\sum_i e^{i(\mathbf{k}'-\mathbf{k})\mathbf{R}_i} &=&
 \frac{(2\pi)^3}{V_0} \delta (\mathbf{k}' - \mathbf{k}) \\
\sum_i e^{i(\mathbf{k}'-\mathbf{k})\mathbf{R}_i} \mathbf{R}_i &=&
\sum_i \nabla_{\mathbf{k}'}
 e^{i(\mathbf{k}'-\mathbf{k})\mathbf{R}_i} = 
\frac{(2\pi)^3}{V_0} \nabla_{\mathbf{k}'}\delta (\mathbf{k}' -
 \mathbf{k})
\lbl{eqA14}
\end{eqnarray}
%For the evaluation of $\nabla_{\mathbf{k}'}\delta (\mathbf{k}' -
%\mathbf{k})$ in (\ref{eqA13}) we use the identity
%\begin{equation}
%f(\mathbf{k}')\nabla_{\mathbf{k}'}\delta (\mathbf{k}' - \mathbf{k})
%= -\delta (\mathbf{k}' - \mathbf{k})
%\nabla_{\mathbf{k}'}f(\mathbf{k}')
%\lbl{eqA15}
%\end{equation}
%with $f(\mathbf{k}')$ resulting from the substitution of (\ref{eqA13})
%into (\ref{eqA12}). Thus:
%\begin{eqnarray}
%&&\frac{\hbar}{im} \int_{V_{\mathrm{BvK}}} d^3r \Psi_{\mathbf{k} a}^*
%\mathbf{p}_{\mathrm{op}} \Psi_{\mathbf{k}' a'} \\
%&=& \frac{(2\pi)^3}{V_0} \delta (\mathbf{k}' - \mathbf{k})
%\left\{ (E_{\mathbf{k} a} - E_{\mathbf{k}' a'}) \int_{V_0}d^3r 
%\Psi_{\mathbf{k} a}^* \mathbf{r} \Psi_{\mathbf{k}' a'}
%- \nabla_{\mathbf{k}} 
%\left[ (E_{\mathbf{k} a} - E_{\mathbf{k}' a'})
%\int_{V_0}d^3r 
%\Psi_{\mathbf{k} a}^* \Psi_{\mathbf{k}' a'} \right] \right\} \\
%&=& \frac{(2\pi)^3}{V_0}\delta (\mathbf{k}' - \mathbf{k})
%\left\{(E_{\mathbf{k} a} - E_{\mathbf{k}' a'}) 
%\int_{V_0}d^3r 
%\Psi_{\mathbf{k} a}^* \mathbf{r} \Psi_{\mathbf{k}' a'}
%-\hbar \mathbf{v}_{\mathbf{k} a} \delta_{aa'}
%- (E_{\mathbf{k} a} - E_{\mathbf{k}' a'}) \nabla_{\mathbf{k}}
%\int_{V_0}d^3r \Psi_{\mathbf{k} a}^* \Psi_{\mathbf{k}' a'}
%\right\}
%\lbl{eqA16}
%\end{eqnarray}
we obtain after some manipulations
\begin{eqnarray}
&&\frac{\hbar}{im} \int_{V_{\mathrm{BvK}}} d^3r \Psi_{\mathbf{k} a}^*
\mathbf{p}_{\mathrm{op}} \Psi_{\mathbf{k}' a'} \nonumber\\
&=& \frac{(2\pi)^3}{V_0}\delta (\mathbf{k}' - \mathbf{k})
\left\{(E_{\mathbf{k} a} - E_{\mathbf{k}' a'}) 
\int_{V_0}d^3r 
\Psi_{\mathbf{k} a}^* \mathbf{r} \Psi_{\mathbf{k}' a'}
-\hbar \mathbf{v}_{\mathbf{k} a} \delta_{aa'}
- (E_{\mathbf{k} a} - E_{\mathbf{k}' a'}) \nabla_{\mathbf{k}}
\int_{V_0}d^3r \Psi_{\mathbf{k} a}^* \Psi_{\mathbf{k}' a'}
\right\}
\lbl{eqA16}
\end{eqnarray}
In all terms, $\mathbf{k} = \mathbf{k}'$ can be directly
substituted due to the $\delta$-function, except in the last one,
where the one must first perform the integration and the
derivation. In the second term, the orthogonality relation
$\int_{V_0}d^3r \Psi_{\mathbf{k}' a}^* \Psi_{\mathbf{k}' a'} =
\delta_{aa'}$ after the substitution $\mathbf{k} = \mathbf{k}'$ has
been used. From this expression we immediately see that in the case of
band crossing, {\it i.e.}~$\mathbf{k} = \mathbf{k}'$,
$E_{\mathbf{k} a} = E_{\mathbf{k}' a'}$, but $a\neq a'$, the
expression vanishes, so the proof is complete. In passing we note
that, if $a=a'$, the expression gives the group velocity as expected. 

%%%%%%%%%%%%%%%%%%%%%%%%%%%%%%%%%%%%%%%%%%%%%%%%%%%%%%%%%%%%%%%%%%%%%%
\section{Current matrix elements in KKR \label{appB}}

In the plane-integration formalism the KKR current matrix elements
read
\begin{equation}
J_{LL'}^{i} = \int_{S_i} d^2r\, R_L(\mathbf{r};E_F) \partial_z
R^*_{L'}(\mathbf{r};E_F)
\lbl{eqCurrentPlane1a}
\end{equation}
where $S_i$ is the surface cut of the atomic cell $i$ with the plane
passing through it.  In the full-potential KKR formalism, the
wavefunctions are expanded in terms of real spherical harmonics as
\begin{equation}
R_L(\mathbf{r}) = \sum_{L_1} \frac{1}{r} R_{L_1 L}(r) Y_{L_1}(\theta,\phi)
\end{equation}
The real spherical harmonics are of the form 
\begin{equation}
Y_L(\theta,\phi) =
  \alpha_L \cdot P_l^{|m|}(\cos\theta) \cdot \mathrm{trg}(m\phi),
\end{equation}
  where 
\begin{eqnarray*}
\alpha_L &=&
  \sqrt{\frac{2l+1}{2\pi}\,\frac{(l-|m|)!}{(l+|m|)!}}, \ m\neq 0 \\
\alpha_L &=&
  \sqrt{\frac{2l+1}{4\pi}}, \ m = 0
\end{eqnarray*}
  $P_l^{|m|}(\cos\theta)$ are the Legendre functions, and
  $\mathrm{trg}(m\phi) = \cos m\phi$, if $m\geq 0$ or $\sin |m|\phi$ ,
  if $m < 0$.  Thus one has to decompose $\partial_z$ into
  $\partial_r$, $\partial_{\theta}$ and $\partial_{\phi}$. The first
  affects only the radial part $R_{L_1 L}(r)$ and is calculated
  numerically; the other two affect only $Y_{L_1}(\theta,\phi)$ and
  are calculated analytically. After some algebra one arrives at the
  result
\begin{eqnarray}
J_{LL'} &=& \int_{r_{\mathrm{min}}}^{r_{\mathrm{max}}} dr\,\frac{1}{r^2}
\nonumber\\
&\times&\sum_{L_2} \alpha_{L_2} \left[ 
\left( r\partial_r R^*_{L_2 L'}(r) - (l_2+1)R^*_{L_2 L'}(r)
\right) u P_{l_2}^{|m_2|}(\cos\theta) \mbox{}+ (l_2+|m_2|) 
P_{l_2-1}^{|m_2|}(\cos\theta) R^*_{L_2 L'}(r)
\right] \nonumber\\
&\times&\sum_{L_1} \alpha_{L_1} P_{l_1}^{|m_1|}(\cos\theta) R_{L_1 L}(r)
\sum_j \int_{\phi_{\mathrm{entry}}^j}^{\phi_{\mathrm{exit}}^j}
\mathrm{trg}(m_1\phi)\mathrm{trg}(m_2\phi)\,d\phi  \,.
\lbl{eqCurrentPlane2}
\end{eqnarray}
Here, $r_{\mathrm{min}}$ and $r_{\mathrm{max}}$ are the radii of the
inscribed and circumscribed circles, respectively, of the convex
polygon, on which the $\phi$-integration is performed, with center the
$z$-projection of the atomic site on the plane;
${\phi_{\mathrm{entry}}^j}$ and ${\phi_{\mathrm{exit}}^j}$ are
respectively angles of entry into and exit from the convex polygon as
the $\phi$-integration is performed.

In the volume-averaging formalism, the KKR current matrix elements
have the form:
\begin{equation}
J^i_{LL'} = \int_{\mathrm{cell}} d^3r 
R^i_L(\mathbf{r}) \partial_z R_{L'}^{i*}(\mathbf{r})
\lbl{eqCurrentVol1}
\end{equation}
These can be computed within the full cell or ASA formalism; here
we shall present both results.

In the full cell approach, the potential is truncated at the
boundary of the Voronoi atomic cell. This is achieved by introducing
the characteristic, or ``shape'', functions
$\Theta(\mathbf{r})$, being equal to unity in the cell
and vanishing outside.\cite{Stefanou90} Their expansion in spherical
harmonics,
\begin{equation}
\Theta(\mathbf{r}) = \sum_L \Theta_L(r) Y_L(\theta,\phi)
\end{equation}
is used in the calculation of the current matrix elements. After some
manipulations we obtain
\begin{eqnarray}
J_{LL'} &=& \int_{\mathrm{WS}} d^3r \, \Theta(\mathbf{r})
R_L(\mathbf{r}) \partial_z R^*_{L'}(\mathbf{r}) \\
 &=& \sum_{L_1}\sum_{L_2}\sum_{L_3} 
\left[ \int dr\, 
\left( R_{L_1 L}(r)\partial_r R^*_{L_2 L'}(r) - \frac{l_2+1}{r}
R_{L_1 L}(r) R^*_{L_2 L'}(r) \right)\Theta_{L_3}(r) \right.
\frac{1}{\alpha_{1,0}} \sum_{L_4} C_{L_1 L_2 L_4} C_{1,0\, L_3 L_4}
\nonumber\\
&&\mbox{}+\left. \int dr\, \frac{1}{r}R_{L_1 L}(r) R^*_{L_2 L'}(r)
  \Theta_{L_3}(r) \frac{\alpha_{l_2 m_2}}{\alpha_{l_2-1, m_2}}
(l_2+|m_2|) C_{L_1 L_3 \, l_2-1,m_2} \right]
\end{eqnarray}
where $C_{L_1L_2L_3}=\int d\Omega Y_{L_1}(\hat{r}) Y_{L_2}(\hat{r})
Y_{L_3}(\hat{r})$ are the Gaunt coefficients and the identity 
$\int d\Omega Y_{L_0}Y_{L_1}Y_{L_2}Y_{L_3} = \sum_{L_4} 
C_{L_1 L_2 L_4} C_{L_0 L_3 L_4}$ has been used.

The ASA result is simpler since it does not involve the shape
functions. The local orbitals have only a spherical part,
\begin{equation}
R_L(\mathbf{r}) = \frac{1}{r} R_l(r)Y_L(\Omega)
\end{equation}
whence the current matrix elements become
\begin{eqnarray}
J_{LL'} &=&\frac{1}{\alpha_{1,0}} C_{LL';1,0} 
\int dr\,R_l(r)\partial_r R^*_{l'}(r)
 - \frac{l'+1}{\alpha_{1,0}} C_{LL';1,0}
\int dr\,\frac{1}{r} R_l(r) R^*_{l'}(r) \nonumber\\
&&\mbox{}+\frac{(l'+|m'|)\alpha_{l'm'}}{\alpha_{l'-1,m'}}
\delta_{l\,l'-1} \delta_{mm'}
\int dr\,\frac{1}{r} R_l(r) R^*_{l'}(r).
\end{eqnarray}
It is implied that the integrals are within the atomic sphere.

\end{widetext}
%%%%%%%%%%%%%%%%%%%%%%%%%%%%%%%%%%%%%%%%%%%%%%%%%%%%%%%%%%%%%%%%%%%%%%


\begin{thebibliography}{99}
\bibitem{footnote1} This can be best understood in a simple
  one-dimensional model of a free electron of wavenumber
  $k_{\mathrm{in}}=\sqrt{E}$ encountering an infinitely long step
  barrier of height $V_0$ where the wavenumber is
  $k_{\mathrm{out}}=\sqrt{E-V_0}$.  The transmission amplitude is
  $t=2k_{\mathrm{in}}/(k_{\mathrm{out}}+k_{\mathrm{in}})$, while the
  transmission probability is $T=4k_{\mathrm{out}}k_{\mathrm{in}}/
  (k_{\mathrm{out}}+k_{\mathrm{in}})^2 =
  |t|^2k_{\mathrm{in}}/k_{\mathrm{out}}$.
\bibitem{footnote2} We remind the reader that Voronoi polyhedra are
  volume-filling convex polyhedra around the atomic sites that
  correspond to a Wigner-Seitz construction, but for arbitrary (not
  necessarily periodic) positions of the atomic sites; moreover, one
  can choose a certain ``weight'' for each atom type in order to give
  it more or less volume fraction.
\end{thebibliography}
\end{document}